\documentclass[12pt]{article}

\usepackage{setspace}
\usepackage[utf8]{inputenc}
\usepackage{amssymb,amsmath,amsfonts}
\usepackage{amsthm}
\usepackage[absolute]{textpos}
\usepackage[
  pdfusetitle,
  bookmarksnumbered,
  hidelinks,
  colorlinks,
  citecolor=blue,
  urlcolor=blue,
  linktoc=page
]{hyperref}
\usepackage{dsfont}
\usepackage{graphicx}
\usepackage{caption}
\usepackage{subcaption}
\usepackage{pgfplots}
\usepackage{slashed}
\usepackage{xspace}
\usepackage{cite}
\pgfplotsset{compat=1.15}

\numberwithin{equation}{section}
\graphicspath{{figures/}}

\usepackage{tikz}
\usetikzlibrary{shapes}

\def\Cm{{\mathcal{C}}}

\def\Fm{{\mathcal{F}}}
\def\Gm{{\mathcal{G}}}
\def\Hm{{\mathcal{H}}}

\def\Mm{{\mathcal{M}}}
\def\Nm{{\mathcal{N}}}
\def\Om{{\mathcal{O}}}

\def\Qm{{\mathcal{Q}}}

\def\Sm{{\mathcal{S}}}

\def\Wm{{\mathcal{W}}}

\def\a{{\alpha}}

\def\ad{{\dot{\alpha}}}

\newcommand\veps{{\varepsilon}}
\newcommand\zb{{\bar z}}
\newcommand\xb{{\bar x}}
\newcommand\hb{{\bar h}}
\newcommand\Dh{{\hat{\Delta}}}
\newcommand\Dp{{\Delta_\phi}}

\newcommand\GFF{\mathrm{GFF}} 
\newcommand\free{\mathrm{free}} 
\newcommand{\mathematica}{\texttt{mathematica}\xspace}
 
\DeclareMathOperator{\dDisc}{dDisc}
\DeclareMathOperator{\Disc}{Disc}
\DeclareMathOperator{\Res}{Res}

\DeclareMathOperator{\INV}{INV}

\DeclareMathOperator{\SINV}{SINV}

\definecolor{nicegreen}{rgb}{0.1,0.6,0.1}

\begin{document}

\begin{textblock}{5}(11,1)
DESY 21-119
\end{textblock}

\title{\textbf{Bootstrapping Monodromy Defects in the \\ Wess-Zumino Model}}
\author{
{\large Aleix Gimenez-Grau, Pedro Liendo} \\ ~ \\[-0.5em]
{\small \emph{DESY Hamburg, Theory Group, Notkestra{\ss}e 85, D-22607 Hamburg, Germany}} \\
{\small Email: \href{mailto:aleix.gimenez@desy.de}{aleix.gimenez@desy.de}, 
\href{mailto:pedro.liendo@desy.de}{pedro.liendo@desy.de}}}
\date{}

\maketitle

\bigskip

\begin{abstract}
	We use analytical bootstrap techniques to study supersymmetric monodromy defects in the critical Wess-Zumino model. In preparation for this result we first study two related systems which are interesting on their own: general monodromy defects (no susy), and the $\veps$--expansion bootstrap for the Wess-Zumino model (no defects). For general monodromy defects, we extend previous work on codimension-two conformal blocks and the Lorentzian inversion formula in order to accommodate parity-odd structures. In the Wess-Zumino model, we bootstrap four-point functions of chiral operators in the $\veps$--expansion, with the goal of obtaining spectral information about the bulk theory. We then proceed to bootstrap two-point functions of chiral operators in the presence of a monodromy defect, and obtain explicit expressions in terms of novel special functions which we analyze in detail. Several of the results presented in this paper are quite general and should be applicable to other setups.
\end{abstract}

\newpage

\tableofcontents

\newpage

\section{Introduction and summary}

Conformal defects are extended objects in conformal field theories that preserve a fraction of the full conformal symmetry. They are important physical observables and their properties should be studied with the same emphasis as the spectrum of local operators. In three dimensions, the critical Ising model has been the subject of intensive research during the past years, and part of this work has focused on its spectrum of defects: conformal boundary conditions were studied using bootstrap techniques in \cite{Liendo:2012hy,Gliozzi:2015qsa}, while the existence of a monodromy defect was proposed in \cite{Billo:2013jda}, and further studied in \cite{Gaiotto:2013nva}.

The motivation behind this work is the study of monodromy defects in the $\mathcal{N}=2$ Wess-Zumino model, which can be considered a supersymmetric counterpart to the standard $3d$ Ising model which preserves four supercharges.\footnote{The $\mathcal{N}=1$ super Ising model can also be formulated as a Wess-Zumino model \cite{Fei:2016sgs}, and has been studied successfully using the numerical bootstrap \cite{Rong:2018okz,Atanasov:2018kqw}.} 
In order to achieve our goal, several intermediate results are necessary, and some of them are interesting on their own right.
In particular, our analysis contains applications valid for non-supersymmetric monodromy defects, for general codimension-two defects and for the Wess-Zumino model without defects. 
The purpose of this detailed introduction is to summarize the paper and provide an outlook of the most relevant results.

Consider a $d$-dimensional Euclidean conformal field theory. 
Whenever there is a complex scalar $\phi(x)$ invariant under $U(1)$ transformations $\phi(x) \to e^{i\alpha} \phi(x)$, a monodromy defect is introduced demanding that the scalar picks a phase when it goes around the origin as follows
\begin{align}
\begin{split}
\label{eq:monodromy-def}
      \phi(r, \theta + 2\pi, \vec y) & = e^{2 \pi i v} \phi(r, \theta, \vec y) \, .
\end{split}
\end{align}
Here $0 \le v < 1$ is a real parameter that characterizes the monodromy, and we are using polar coordinates $(r,\theta)$ in the plane orthogonal to the defect.
The critical Ising model provides the simplest example: since the global symmetry is $\mathbb{Z}_2$, there exists a monodromy defect with $v = 1/2$.
This defect was studied in \cite{Billo:2013jda,Gaiotto:2013nva} using Monte-Carlo simulations, $\veps$--expansion calculations and numerical bootstrap (see also \cite{Yamaguchi:2016pbj}).
For the case of the $O(N)$ models, there exist monodromy defects with general $v$, which were studied in the $\veps$--expansion in \cite{Soderberg:2017oaa}, and recently the very systematic study of \cite{Giombi:2021uae} has extended these results and obtained new ones in the large-$N$ limit.\footnote{
The monodromy defect geometry is reminiscent of two intersecting boundaries at an angle $\theta = 2\pi v$, although the later setup breaks more symmetry \cite{Antunes:2021qpy}.}

In the present work, an important observable we consider are two-point correlation functions of scalar fields in the presence of a monodromy defect.
Since the monodromy partly breaks conformal symmetry, the two-point function depends on two conformal cross ratios $x$ and $\xb$, to be defined in \eqref{eq:def-cross-ratios}.
As a result, the correlator reads
\begin{align}
\label{eq:two-pt-def}
 \langle \phi(x_1) \bar \phi(x_2) \rangle
 = \frac{\Gm(x,\xb)}{(r_1 r_2)^{\Dp}} \, .
\end{align}
Analogously to four-point functions in homogeneous CFT, the correlator $\Gm(x,\xb)$ captures an infinite amount of CFT data thanks to the Operator Product Expansion (OPE).
In the presence of a defect, two different OPEs are possible, one as a sum of bulk operators, the other in terms of operators localized on the defect \cite{Billo:2016cpy}.
For two-point functions, these OPEs give two conformal block decompositions which must be equal:\footnote{Here and in the rest of the paper we use the shorthand notation $\mu_{\Dh,s} = |b_{\phi\widehat\Om}|^2$ and $c_{\Delta,\ell} = a_{\Om} \lambda_{\phi\bar\phi\Om}$, where $b_{\phi\widehat\Om}$, $a_{\Om}$, $\lambda_{\phi\bar\phi\Om}$ are OPE coefficients defined in the main text.}
\begin{align}
\label{eq:crossing-eq}
 \Gm(x, \xb)
 = \sum_{\Dh,s} \mu_{\Dh,s} \hat f_{\Dh,s}(x, \xb)
 = \left( \frac{\sqrt{x \xb}}{(1-x)(1-\xb)}\right)^\Dp
 \sum_{\Delta,\ell} c_{\Delta,\ell} f_{\Delta,\ell}(x, \xb) \, .
\end{align}
In this paper, we follow the bootstrap philosophy which uses the crossing equation \eqref{eq:crossing-eq} as the starting point.
Indeed, we will see that in favorable situations, \eqref{eq:crossing-eq} together with basic structural properties of the bulk theory and mild physical assumptions, can be used to fully determine the correlator $\Gm(x,\xb)$. 
In the case of conformal boundaries, this approach has been successfully carried out in a number of interesting examples \cite{Liendo:2012hy,Bissi:2018mcq,Mazac:2018biw,Kaviraj:2018tfd,Gimenez-Grau:2020jvf,Dey:2020jlc}.

The main technical tool we will use to solve crossing analytically is the so-called Lorentzian inversion formula (LIF).
The original LIF was derived for four-point correlation functions in CFTs without defects \cite{Caron-Huot:2017vep}.
In the case of two-point functions in defect CFT, there exist two inversion formulas, one for each of the OPE channels in the crossing equation \eqref{eq:crossing-eq}.
These formulas were obtained in \cite{Lemos:2017vnx,Liendo:2019jpu} and were already used to study the $\mathbb Z_2$ Ising monodromy defect. In this work, we continue with this program and use the LIF to solve more general monodromy defects in the $\veps$--expansion. 

We start in section \ref{sec:wf-monodromy} with the Wilson-Fisher (WF) fixed point with global $O(2N)$ symmetry.
This model is described in $d = 4-\veps$ dimensions by the non-trivial fixed point of the following Lagrangian
\begin{align}
\begin{split}
\label{eq:wf-lagrangian}
 L_{\text{WF}}
   = \frac12 (\partial_\mu \phi_i)^2
   + \frac{\lambda}{4!} (\phi_i \phi_i)^2 \, , \qquad
 i = 1, \ldots, 2N \, .
\end{split}
\end{align}
We define the complex scalar $\phi = \phi_1 + i \phi_2$ and impose a monodromy $v$ under rotations \eqref{eq:monodromy-def}.
Since this model is weakly coupled for $0 < \veps \ll 1$, one can use the Lagrangian description to compute CFT data using Feynman diagrams \cite{Gaiotto:2013nva,Soderberg:2017oaa,Giombi:2021uae}.
However, this is not the approach we follow on this work. Although still perturbative in nature, our analysis relies solely on modern analytical bootstrap techniques. The bootstrap has several advantages which allow us to present improvements on previous results.
On the one hand, we obtain closed-form expressions for the correlation function $\Gm(x,\xb)$ to order $O(\veps)$, which allows us to extract previously unknown bulk CFT data in an efficient way.
On the other hand, we show that the correlator is an analytic function of the monodromy $v$, and the transformation $v \to v+1$ has the interpretation of a change of boundary condition for low-lying defect operators.
We also clarify subtleties related to codimension-two defects that had not appeared in the literature.
In particular, we obtain conformal blocks for odd-spin bulk operators, which are related to the existence of parity-odd one-point tensor structures when the codimension is two.
In order to accommodate these operators, we also have to extend the bulk-to-defect Lorentzian inversion formula \cite{Lemos:2017vnx}.
These results not only are applicable to monodromy defects, but to any type of codimension-two defect.

Having used the Wilson-Fisher model as a testing ground for our techniques, we move on to the Wess-Zumino (WZ) model, which is the simplest superconformal model preserving four supercharges.
This model consists of a complex scalar $\phi(x)$ and a two-component complex fermion $\psi(x)$.
The allowed interactions are fully fixed by supersymmetry, so the action depends on a single coupling constant $g$:
\begin{align}
\label{eq:wz-lagrangian}
 L_{\text{WZ}}
 = (\partial_\mu \bar \phi) (\partial_\mu \phi)
 + \psi^\dag \bar \sigma^\mu \partial_\mu \psi
 + \frac g2 (\psi \psi \phi + \psi^\dag \psi^\dag \bar \phi)
 + \frac {g^2}{4} (\phi \bar \phi)^2 \, .
\end{align}
Similarly to the WF case, this model has a fixed point in $d = 4-\veps$ dimensions that can be studied in diagrammatic perturbation theory.\footnote{See \cite{Fei:2016sgs} for a nice summary and introduction to the literature.}
Compared to the Wilson-Fisher fixed point, which has gotten a lot of attention from the bootstrap community  \cite{Gopakumar:2016wkt,Gopakumar:2016cpb,Dey:2016mcs,Dey:2017oim,Alday:2017zzv,Henriksson:2018myn,Henriksson:2020fqi,Henriksson:2021lwn}, the literature on the Wess-Zumino model using modern conformal bootstrap is much scarcer, the most notable exceptions being \cite{Bobev:2015jxa,Bobev:2015vsa}.

In section \ref{sec:wess-zumino-bulk} we take a small detour in order to fill this gap.
In this section we forget momentarily about defects, and we start by modifying the original LIF \cite{Caron-Huot:2017vep} into a formula that directly extracts OPE coefficients of exchanged superconformal primaries.
The main virtues of this formula are that it unmixes the contributions of nearly-degenerate operators, and that it applies to general superconformal theories with four supercharges in any number of dimensions.
With this newly developed machinery, we carry out the bootstrap program for bulk four-point functions of chiral operators and extract bulk CFT data to leading order in $\veps$.
This is the simplest application of our formalism, and we hope to present a more detailed treatment of the Wess-Zumino model using LIF technology elsewhere. 

In section \ref{sec:wess-zumino-defect} we put all the pieces together and study monodromy defects in the Wess-Zumino model.
We start by reviewing the relevant superconformal blocks \cite{Gimenez-Grau:2020jvf}, and then move on to use the input of section \ref{sec:wess-zumino-bulk} and the LIF to bootstrap two-point functions of chiral fields.
The final result can be written in a compact form in terms of a class of one- and two-variable special functions which are defined by their series expansions. Because these functions might be relevant for future bootstrap calculations, we study some of their analytic properties in more detail. In particular, we explain how to extract their behavior around $x,\xb \sim 1$ given their series expansions around $x,\xb \sim 0$. This amounts to extracting both bulk and defect CFT data to leading order in  $\veps$, which was one of the original goals of this work.

\section{Wilson-Fisher: Monodromy defects}
\label{sec:wf-monodromy}

In this section we study monodromy defects in the Wilson-Fisher fixed point, previous work on this subject include~\cite{Soderberg:2017oaa,Giombi:2021uae}.\footnote{See also \cite{Bianchi:2021snj,Dowker:2021gqj} for other works using methods slightly different to ours.} Here we present some small improvements by obtaining the full correlation function at order $O(\veps)$ and extracting the bulk CFT data.
This model, interesting on its own, is also a good testing ground for our techniques, which we will later apply to the Wess-Zumino model in section \ref{sec:wess-zumino-defect}.

We start this section studying kinematics of codimension-two defects in $d$-dimensional Euclidean spacetime.
Even though kinematics of defect CFTs are well understood in general \cite{Billo:2016cpy}, the codimension-two case turns out to be subtle.
In particular, we obtain bulk conformal blocks for odd-spin operators, which have not appeared in the literature before.
Furthermore, we extend the bulk-to-defect inversion formula of \cite{Lemos:2017vnx}, in order to accommodate odd-spin operators for generic codimension-two defects.
We end the section by bootstrapping two-point functions of bulk scalars $\langle \phi(x_1) \bar \phi(x_2) \rangle$ in the presence of monodromy defects, first for free theories, and then for the more interesting case of the Wilson-Fisher fixed point. 

\subsection{Conformal cross ratios}

As anticipated in the introduction, the two-point function of scalars in the presence of a defect depends on a function of two conformal cross ratios
\begin{align}
\label{eq:two-pt-def-2}
 \langle \phi(x_1) \bar \phi(x_2) \rangle
 = \frac{\Gm(x,\xb)}{|x_1^\bot|^{\Dp} |x_2^\bot|^{\Dp}} \, .
\end{align}
In this work, we use the same cross ratios as \cite{Lemos:2017vnx}, which are defined by\footnote{Our cross-ratios are related to the ones in \cite{Giombi:2021uae} as $e^{i\theta} = \sqrt{x/\xb}$ and $\xi = (1-\sqrt{x\xb})^2 / (4 \sqrt{x\xb})$.}
\begin{align}
 \sqrt{x \xb} + \frac{1}{\sqrt{x \xb}} 
 = \frac{|x_{12}^{\|}|^2 + |x_1^\bot|^2 + |x_2^\bot|^2}{|x_1^\bot| |x_2^\bot|} 
 \, , \qquad
 \sqrt{\frac{x}{\xb}} + \sqrt{\frac{\xb}{x}}
 = \frac{2 x_1^\bot \cdot x_2^\bot}{|x_1^\bot| |x_2^\bot|} 
 \, .
\label{eq:def-cross-ratios}
\end{align}
Here we are assuming a flat defect, with $x^{\|}$ directions parallel to the defect and $x^\bot$ orthogonal directions.

In order to give a geometric interpretation of the cross ratios, it is convenient to use a conformal transformation to go to a simpler frame.
In the frame of interest, the defect sits at the origin, the two operators $\phi(x_1)$, $\bar\phi(x_2)$ lie on a plane orthogonal to the defect, and $\bar \phi(x_2)$ is set at one.
The position of $\phi(x_1)$ is unfixed and depends on two coordinates, which are precisely the two cross ratios in \eqref{eq:two-pt-def-2}.
In Euclidean signature, it is convenient to parametrize the position of $\phi(x_1)$ with complex conjugate coordinates $x$ and $\xb = x^*$, namely:
\begin{align}
\label{eq:cross-ratios-plane}
 x_1 = \left( \tfrac12 (x + \xb), \tfrac{1}{2i}(x-\xb), \vec y \right), \qquad
 x_2 = \left( 1, 0, \vec y \right).
\end{align}
Here $\vec y$ parametrizes the directions parallel to the defect.
Continuing the CFT to Lorentzian signature, one sees that the two cross ratios $x$, $\xb$ become real and independent.
Because of their interpretation as coordinates in a plane and their reality conditions, the defect CFT cross ratios $x,\xb$ are close analogs of the four-point cross-ratios $z,\zb$ which are familiar in homogeneous CFT.

\subsection{Conformal blocks}

The two point function $\Gm(x,\xb)$ admits two different expansions, the defect-channel expansion and the bulk OPE expansion, see \eqref{eq:crossing-eq}.
These expansions are formulated in terms of the conformal blocks that we now study.
The discussion that follows is always restricted to codimension-two defects.

\subsubsection{Defect channel}

The defect-channel expansion expresses a bulk field as an infinite sum of defect fields.
In the coordinates $x$, $\xb$ of equation \eqref{eq:cross-ratios-plane}, the defect sits at $x=\xb=0$, while the operator $\bar \phi(x_2)$ sits at $x = \xb = 1$.
The defect OPE limit dominates when $\phi(x_1)$ approaches the defect, namely when $x\xb \to 0$ keeping $x/\xb$ fixed.
To leading order in $x\xb$ and to all orders in $x/\xb$, we normalize the defect expansion as
\begin{align}
\label{eq:leading-def-ope}
 \phi(x, \xb, \vec y) 
 \sim \sum_{\widehat \Om} b_{\phi \widehat \Om} 
      \left( \frac{\xb}{x} \right)^s 
      ( x \xb)^{(\widehat \Delta-\Dp)/2} 
      \left[ \widehat \Om(\vec y) + O(x\xb) \right] \, .
\end{align}
Inserting \eqref{eq:leading-def-ope} in the two-point function and comparing with the defect expansion \eqref{eq:crossing-eq} gives the leading behavior of defect blocks 
\begin{align}
 \label{eq:bdy-cond}
 \hat f_{\hat\Delta,s}(x, \xb) \sim x^{(\Dh-s)/2} \xb^{(\Dh+s)/2} + O(x\xb) \, .
\end{align}
The full cross-ratio dependence of the conformal block $\hat f_{\hat\Delta,s}(x,\xb) = (\xb/x)^s g_{\hat\Delta}(x\xb)$ can be determined from the Casimir equation derived in \cite{Billo:2016cpy}, namely
\begin{align}
 \left[
 (1-y) y^2 \partial_y^2 
 - \frac{1}{2} y (d y+d-4) \partial_y
 - \frac{1}{4} \Dh (\Dh-d+2) (1-y) \right] g(y)
 = 0 \, .
\end{align}
This equation has two hypergeometric solutions, and the one with the correct boundary conditions \eqref{eq:bdy-cond} leads to the final form of defect-channel conformal blocks:
\begin{align}
\label{eq:cod2-def-block}
\begin{split}
 \hat f_{\hat\Delta,s}(x, \xb) 
 & = x^{(\Dh-s)/2} \xb^{(\Dh+s)/2} 
   {}_2F_1 \big( \Dh,d/2-1;\Dh + 2 - d/2; x \xb \big) \,.
\end{split}
\end{align}

\paragraph{Monodromy defects:}
Even though this section applies to arbitrary codimension-two defects, let us return momentarily to monodromy defects.
Since we work in Euclidean signature, the cross ratios are complex conjugates of each other $x^* = \xb$, and moving $\phi(x_1)$ around the defect corresponds to analytically continuing $x$ ($\xb$) around the origin counterclockwise (clockwise).
Together with \eqref{eq:monodromy-def}, we conclude that our correlation function must satisfy the boundary condition
\begin{align}
 \label{eq:twist-def}
 \Gm(x^\circlearrowleft, \xb^\circlearrowright) = e^{+2\pi i v} \Gm(x, \xb) \, .
\end{align}
The monodromy \eqref{eq:twist-def} combined with the form of the defect block \eqref{eq:cod2-def-block} requires the defect spectrum to consists of non-integer transverse spins:
\begin{align}
\label{eq:monodromy-defect-spins}
 s = -v + n \quad \text{for} \quad n \in \mathbb Z \, .
\end{align}
This observation will be important in modifying the Lorentzian inversion formula in section \ref{sec:inv-form}, and in the study of monodromy defects starting in section \ref{sec:gff-non-susy}.

\subsubsection{Bulk channel}
\label{sec:bulk-blocks}

Let us turn to the bulk-channel expansion, where the product $\phi(x_1) \bar \phi(x_2)$ is expanded in an infinite sum of bulk operators by means of the usual operator product expansion (OPE).
In the frame of equation \eqref{eq:cross-ratios-plane}, the second operator is located at $x = \xb = 1$, so the bulk-channel expansion dominates in the regime $(1-x)(1-\xb) \to 0$.
Since $\phi$ and $\bar \phi$ are unequal operators, the bulk OPE consists of both even and odd spin operators.
As is customary, we use index free notation $\Om^{(\ell)}(x, u) = \Om^{\mu_1 \ldots \mu_\ell}(x) u_{\mu_1} \ldots u_{\mu_\ell}$ and assume the following normalization for the OPE\footnote{The awkward factor $2^{\ell/2}$ leads to four-point blocks normalized as $g_{\Delta,\ell}(z,\zb) \sim z^{(\Delta-\ell)/2} \zb^{(\Delta+\ell)/2}$ in the lightcone limit.}
\begin{align}
\begin{split}
 & \phi(x_1) \bar \phi(x_2) 
 \sim \sum_{\Om^{(\ell)}} \lambda_{12\Om} \, 2^{\ell/2}
      \frac{\Om^{(\ell)}(x_2, x_{12})}{x_{12}^{\Delta_1+\Delta_2-\Delta+\ell}} 
      + \ldots \, ,
\end{split}
\end{align}
where we keep the leading order in the bulk OPE limit $x_{12}^2 \to 0$.
For general defects, only even-spin operators can have one-point functions \cite{Billo:2016cpy}.
However, a peculiarity of codimension-two defects is that odd-spin operators can also have one-point functions:
\begin{align}
\label{eq:one-pt-funcs}
\begin{split}
 \ell \text{ even:} \qquad
 & \langle \Om^{(\ell)}(x, u) \rangle
 = \frac{2^{\ell/2} a_\Om}{|x^i|^{\Delta}} 
   \left( \frac{(x^i u^i)^2}{|x^i|^{2}} - u^i u^i \right)^{\ell/2} \, , \\
 \ell \text{ odd:} \qquad
 & \langle \Om^{(\ell)}(x,u) \rangle
 = - \frac{i \, 2^{\ell/2} a_\Om \, \veps_{ij} u^i u^j}{|x^i|^{\Delta+1}} 
   \left( \frac{(x^i u^i)^2}{|x^i|^{2}} - u^i u^i \right)^{(\ell-1)/2} \, .
\end{split}
\end{align}
Here $i,j=1,2$ are indices in the two directions orthogonal to the defect, and $\veps_{ij}$ is the two-index antisymmetric tensor, which is an allowed tensor structure for codimension-two defects.
Combining the bulk OPE with the form of the one-point function gives the leading order behavior of blocks with even and odd spin:
\begin{align}
 \lim_{x,\xb \to 1} f_{\Delta,\ell}(x, \xb)
 = \big[ (1-x)(1-\xb) \big]^{(\Delta-\ell)/2} (x - \xb)^\ell \, .
\end{align}
It is perhaps surprising that odd-spin bulk blocks are antisymmetric under $x \leftrightarrow \xb$, but it is a direct consequence of the existence of parity-odd one-point functions \eqref{eq:one-pt-funcs}.
It is also interesting to consider the normalization of bulk blocks in the lightcone limit
\begin{align}
\label{eq:lightcone-asymptotics}
 f_{\Delta,\ell}(x, \xb) 
 = \left\{
 \begin{array}{ll}
  (1-x)^{(\Delta-\ell)/2} (1-\xb)^{(\Delta+\ell)/2} & 
 \qquad 0 < 1-x \ll 1-\xb \ll 1 \, , \\
 (-1)^\ell (1-\xb)^{(\Delta-\ell)/2} (1-x)^{(\Delta+\ell)/2} & 
 \qquad 0 < 1-\xb \ll 1-x \ll 1 \, .
 \end{array}
 \right.
\end{align}
As before, the full dependence of $f_{\Delta,\ell}$ on the cross-ratios can be obtained by solving the Casimir differential equation, which has been worked out in \cite{Billo:2016cpy,Isachenkov:2018pef}.
We are interested in the codimension-two case, when the differential operator in $x,\xb$ coordinates reads
\begin{align}
\label{eq:cas-eq}
\begin{split}
 & \left( 
    D_x
    + D_\xb 
    + (d-2) \frac{(1-x)(1-\xb)}{1 - x \xb} 
      \big( x \partial_x + \xb \partial_\xb \big)
    - \frac{1}{2} c_2
 \right) f_{\Delta,\ell}(x,\xb)
 = 0 \, , \\
 & D_x
 = (1-x)^2 x \partial^2_x
 + (1-x)^2 \partial_x \, ,
\end{split}
\end{align}
and the Casimir eigenvalue is $c_2 = \Delta(\Delta-d) + \ell (\ell + d -2)$.
The similarity of \eqref{eq:cas-eq} with the Dolan and Osborn differential operator \cite{Dolan:2003hv,Dolan:2011dv} is apparent.
Indeed, it was originally pointed out in \cite{Billo:2016cpy} that in terms of $z,\zb$ coordinates
\begin{align}
\label{eq:map-xz}
 x = 1-z \, , \qquad
 \xb = (1 - \zb)^{-1} \, ,
\end{align}
the two differential operators are the same.
By comparing the lightcone limit of the defect block \eqref{eq:lightcone-asymptotics} with the lightcone limit of four-point blocks, we obtain the precise mapping
\begin{align}
\label{eq:map-4pt-def-blocks}
 f_{\Delta,\ell}(x, \xb)
 = (-1)^{-(\Delta+\ell)/2} 
   g_{\Delta,\ell}\left(1-x, \frac{\xb-1}{\xb}\right).
\end{align}
Our discussion makes it clear that this relation is valid both for even- and odd-spin bulk operators.
In the four-dimensional case, which is relevant for the present work, simple closed-form expressions for the four-point blocks are known \cite{Dolan:2000ut}, which in the defect case map to
\begin{align}
\label{eq:bulk-4d-blocks}
\begin{split}
 f_{\Delta,\ell}(x, \xb)
 & = \frac{(1-x)(1-\xb)}{1 - x\xb} \Big( 
     k_{\Delta-\ell-2}^{0,0}(1-x) k_{\Delta+\ell}^{0,0}(1-\xb)
   + (-1)^\ell \big( x \leftrightarrow \xb \big)
 \Big) \, , \\
 k^{r,s}_\beta(x)
 & = x^{\beta/2} 
    {}_2F_1 \left( \frac{\beta-r}{2}, \frac{\beta+s}{2}, \beta, x \right) \, .
\end{split}
\end{align}
It is easy to check that this is normalized according to \eqref{eq:lightcone-asymptotics}.
For general space-time dimensions $d$, one makes an ansatz of the form \cite{Simmons-Duffin:2016wlq}
\begin{align}
\label{eq:ligh-exp-bos-blocks}
 f_{\Delta,\ell}(x, \xb)
 = \sum_{n=0}^\infty \sum_{j=-n}^n A_{n,j}(\Delta,\ell) 
    (1-x)^{(\Delta-\ell)/2+n} k_{\Delta+\ell+2j}^{0,0}(1-\xb) \, ,
\end{align}
and fixes the coefficients recursively with the Casimir equation \eqref{eq:cas-eq}.
This process can be implemented efficiently using a computer.
For the sake of clarity, we present some low-lying coefficients:
\begin{align}
 A_{0,0} (\Delta,\ell) = 1 \, , \qquad
 A_{1,0} (\Delta,\ell) = \frac{\Delta -\ell}{4} \, , \qquad
 A_{1,-1}(\Delta,\ell) = -\frac{(d-2) \ell}{2 \ell + d - 4} \, .
\end{align}

\subsection{Bulk-to-defect inversion formula}
\label{sec:inv-form}

The Lorentzian Inversion Formula (LIF) \cite{Caron-Huot:2017vep,Simmons-Duffin:2017nub} is a central tool for the analytic bootstrap program.
In the presence of defects, one can consider a bulk-to-defect LIF \cite{Lemos:2017vnx} and a defect-to-bulk LIF \cite{Liendo:2019jpu}.
The bulk-to-defect LIF is of particular importance in this work, as will become clear in subsequent sections.
For codimension-two defects, we need a small extension of the formula presented in \cite{Lemos:2017vnx} which we outline below, and we refer the reader to \cite{Lemos:2017vnx} for further details.

\subsubsection{Derivation}

The LIF of \cite{Lemos:2017vnx} was derived assuming that the correlator $\Gm(x,\xb)$ is a symmetric function of $x,\xb$, which is true when the external scalars are identical and the theory preserves parity. In our setup, the bulk expansion generically contains even- and odd-spin blocks, which are symmetric and antisymmetric respectively, so the full correlator has no definite symmetry. Furthermore, our derivation is valid for non-integer values of $s$, which is the relevant situation for monodromy defects.

The central object of this discussion is the function $\mu(\Delta,s)$, which encodes dimensions of defect operators as poles and their OPE coefficients as residues:
\begin{align}
\label{eq:residue}
\mu_{\Dh_*, s} \equiv b^2_{\Dh_*, s} = - \text{Res}_{\Dh = \Dh_*} \mu(\Dh, s) \, .
\end{align}
Let us introduce coordinates $x = rw$ and $\xb = r/w$, which in Euclidean signature correspond to a radial coordinate $r$ and a phase $w$.
The conformal block \eqref{eq:cod2-def-block} can be decomposed as $\hat f_{\Dh,s}(r,w) = w^{-s} \hat f_{\Dh}(r)$, and the correlation function admits a partial wave expansion
\begin{align}
\label{eq:partial-wave-decomp}
 \Gm(r, w) 
 = \sum_{s} \int_{p/2-i\infty}^{p/2+i\infty} 
   \frac{d\Dh}{2\pi i} \mu(\Dh, s) w^{-s} \Psi_\Dh(r) \, , \quad
 \Psi_{\Dh}(r)
 \equiv \frac{1}{2} \left( 
    \hat f_{\Dh} 
    + \frac{K_{p-\Dh}}{K_{\Dh}} \hat f_{p-\Dh} \right) ,
\end{align}
where the sum runs for all $-\infty < s < \infty$ and we introduced $K_{\Dh} = \Gamma(\Dh) / \Gamma(\Dh - p/2)$ and $p = d-2$.
When the partial wave $\Psi_{\Dh}(r)$ has dimension $\Dh = p/2 + i \nu$ it obeys an orthonormality relation \cite{Lemos:2017vnx}:
\begin{align}
\begin{split}
 & \int_0^1 dr \frac{(1-r^2)^{d-2}}{r^{d-1}} \Psi_{\Dh_1}(r) \Psi_{\Dh_2}(r)
 = \frac{\pi}{2} \frac{K_{p-\Dh_2}}{K_{\Dh_1}} 
   [ \delta(\nu_1 - \nu_2) + \delta(\nu_1 + \nu_2) ] \, .
\end{split}
\end{align}
Furthermore, we assume the defect spectrum is such that the transverse spins are integer separated $s_1 - s_2 \in \mathbb{Z}$.
In this case, we have the orthonormality relation
\begin{align}
  & \oint \frac{dw}{2\pi i w} w^{s_1-s_2} = \delta_{s_1,s_2} \, ,
\end{align}
where the integral is along the unit circle $|w| = 1$.
Combining the partial wave decomposition \eqref{eq:partial-wave-decomp} with the orthonormality of our basis, one readily obtains the Euclidean inversion formula:
\begin{align}
 \mu(\Dh, s)
 = \frac{2 K_\Dh}{K_{p-\Dh}}
 \oint \frac{dw}{2\pi i w} w^{s}
 \int_0^1 dr \frac{(1-r^2)^{d-2}}{r^{d-1}} \Psi_\Dh(r) \Gm(r, w) \, .
\end{align}
Let us stress that this formula is only valid for physical values of the transverse spin $s$.
Now we would like to deform the integration contour of $w$ into Lorentzian kinematics, leading to a formula analytic in $s$.
However, in order to deform the contour safely, one needs the asymptotic behavior of $\Gm(r, w)$ for large and small $w$:
\begin{align}
\label{eq:regge-behavior}
 \Gm(r, w) \lesssim w^{-s^*_+} \quad \text{as} \quad w \to 0 \, , \qquad
 \Gm(r, w) \lesssim w^{ s^*_-} \quad \text{as} \quad w \to \infty \, .
\end{align}
Then we conclude that for $s > s^*_+$ we can contract the contour towards the origin picking up a discontinuity around the cut $w \in [0, r]$. 
Similarly, for  $s < -s^*_-$ we blow up the contour to infinity, picking a discontinuity around the cut $w \in [1/r, \infty]$.
We then rewrite the resulting integral in terms of $x,\xb$, and keep only poles in $\mu(\Dh,s)$ corresponding to the exchanged operator and not its shadow.
After the dust settles, we obtain the bulk-to-defect Lorentzian inversion formula in its final form:
\begin{align}
\label{eq:inv-formula}
 \mu(\Dh, s) 
 = \begin{cases}
  \int_0^1 dx \int_1^{1/x} d\xb \, I_{\Dh,s}(x, \xb) \Disc_\xb \Gm(x, \xb) 
  \quad \text{for} \quad s>s^*_+ \\
  \int_0^1 d\xb \int_1^{1/\xb} dx \, I_{\Dh,s}(x, \xb) \Disc_x \Gm(x, \xb) 
  \quad \text{for} \quad s<-s^*_- \\
 \end{cases} .
\end{align}
In the above formula, the integration kernel and discontinuities are given by:
\begin{align}
\label{eq:details-lif}
\begin{split}
 & I_{\Dh, s}(x, \xb) 
 = \frac{1}{4 \pi i}
   \, x^{-\frac{\Dh-s+2}{2}} \xb^{-\frac{\Dh+s+2}{2}} (1 - x \xb) 
   {}_2F_1 \bigg( \! {\begin{array}{c c}
    {1-\Dh, 2-d/2} \\
    {d/2-\Dh}
    \end{array}; x \xb} \bigg) \, , \\
 & \Disc_x \Gm(x, \xb)
 = \Gm(x + i 0, \xb) - \Gm(x - i0, \xb) \, , \\
 & \Disc_\xb \Gm(x, \xb)
 = \Gm(x, \xb + i 0) - \Gm(x, \xb - i0) \, .
\end{split}
\end{align}
This is equal to the inversion formula obtained in \cite{Lemos:2017vnx} for $s>s^*_+$, but one has to exchange the role of $x \leftrightarrow \xb$ to obtain the defect CFT data for for $s<-s^*_-$.
The difference arises because \cite{Lemos:2017vnx} assumed that the correlator $\Gm(x,\xb)$ is a symmetric function of $x$, $\xb$, which is true for defects of codimension greater than two and for codimension-two defects without parity-odd operators.
Instead, here we focus on codimension two and allow $\Gm(x,\xb)$ to have no definite symmetry.
As we will see in section \ref{sec:wess-zumino-defect}, this extension of the original LIF is necessary for applications in the Wess-Zumino model.

Let us also mention that for the particular case when $\Gm(x,\xb)$ is symmetric, the LIF can be simplified.
Indeed, for symmetric correlators equation \eqref{eq:regge-behavior} implies $s_-^* = s_+^* \equiv s^*$ and the two contributions in the inversion formula can be combined:
\begin{align}
\label{eq:inv-formula-even}
 \mu(\Dh, s) 
 = \int_0^1 dx \int_1^{1/x} d\xb \, I_{\Dh,|s|}(x, \xb) \Disc_\xb \Gm(x, \xb) 
  \;\;  \text{for} \;\; |s|>s^* \;\; \text{and} \;\; \Gm(x,\xb) = \Gm(\xb,x) \, .
\end{align}
The advantage is that now one recovers the positive and negative transverse-spin trajectories at the same time.

\subsubsection{Applications}
\label{sec:lif-applications}

Let us also briefly discuss how to use the inversion formula in practice.
The inversion formula uses the discontinuity across branch cuts that start at $x,\xb=1$.
It is thus possible to compute this discontinuity term by term using the bulk-channel expansion, which is an expansion in powers of $(1-x)$, $(1-\xb)$.
If follows from section \ref{sec:bulk-blocks} that bulk blocks have the structure
\begin{align}
\label{eq:red-blocks}
 f_{\Delta,\ell}(x,\xb)
 & = \big[ (1-x)(1-\xb) \big]^{(\Delta-\ell)/2} \tilde f_{\Delta,\ell}(x, \xb) \, .
\end{align}
Here the prefactor is possibly non-analytic around $x,\xb = 1$, while $\tilde f_{\Delta,\ell}$ is analytic at $x,\xb = 1$. 
Equivalently, $\tilde f_{\Delta,\ell}$ admits a convergent power series in integer powers of $1-x$ and $1-\xb$:
\begin{align}
 \tilde f_{\Delta,\ell}(x, \xb)
 = \sum_{n,m \ge 0} k_{n,m} (1-x)^n (1-\xb)^m \, .
\end{align}
As a result, the discontinuity picks only the contribution from the prefactor in \eqref{eq:red-blocks}, so focusing on $\Disc_\xb$ for concreteness
\begin{align}
\label{eq:compute-disc}
\begin{split}
 \Disc_\xb \Gm(x,\xb)
&= \Disc_\xb \left( \frac{\sqrt{x \xb}}{(1-x)(1-\xb)} \right)^{\Dp} 
   \sum_{\Delta,\ell} c_\Om f_{\Delta,\ell}(x,\xb) \\
&= \left( \frac{\sqrt{x \xb}}{(1-x)} \right)^{\Dp} 
   \sum_{\Delta,\ell} c_\Om \tilde f_{\Delta,\ell}(x,\xb)
   \Disc_\xb (1-\xb)^{\frac{\Delta-\ell}{2} - \Dp} \, .
\end{split}
\end{align}
It should be mentioned here that there are two ways for $\Disc_\xb (1-\xb)^\a \ne 0$: if $\a$ is non-integer, or if $\alpha = -n$ is a negative integer.
In these two cases the discontinuity reads
\begin{align}
\label{eq:disc-cases}
\begin{split}
 & \Disc_\xb (1-\xb)^\a = 2i \sin (\pi \a) \, (\xb-1)^\a
 \quad \text{for} \quad \alpha \notin \mathbb{N} \, , \\
 & \Disc_\xb \frac{1}{(1-\xb)^n} = 2 \pi i \frac{(-1)^n}{(n-1)!} \delta^{(n-1)}(1-\xb)
 \quad \text{for} \quad n \in \mathbb{N}_+ \, .
\end{split}
\end{align}
The first formula follows straightforwardly from the definition of discontinuity \eqref{eq:details-lif}, while the second can be justified by integrating against a test function, see for example (3.7) in \cite{Bissi:2019kkx}.

All in all, comparing \eqref{eq:compute-disc} and \eqref{eq:disc-cases}, it is clear that only two classes of bulk operators contribute to the inversion formula:
\begin{enumerate}
 \item Operators below the double-twist dimension $\Delta < 2 \Dp + \ell$. 
 The most important example of this kind is that bulk identity $\Delta = \ell = 0$, which is present in any CFT. 
 This contribution will be studied in detail in section \ref{sec:gff-non-susy}.
 Another example are single-trace operators in large-$N$ CFTs \cite{Barrat:2021yvp}, but they play no role in the present paper. 
 \item Double-twist operators with anomalous dimension $\Delta = 2 \Dp + \ell + 2n + \gamma$. These operators are the ones that will contribute in our study of the Wilson-Fisher and Wess-Zumino models in subsequent sections.
\end{enumerate}
Summarizing, the LIF kills bulk operators with exact double-twist dimension $\Delta = 2\Dp + \ell + 2n$. 
This is ultimately the reason why the LIF is so powerful.

\subsection{GFF monodromy defect}
\label{sec:gff-non-susy}

Having developed the necessary techniques, we are ready to study monodromy defects using analytic bootstrap.
We start with a generalized free field (GFF) $\phi(x)$  of dimension $\Dp$.
It is well known that the bulk spectrum of GFF consists of the identity and double-twist operators $\Delta_{\ell,n} = 2\Dp + \ell + 2n$, and we just discussed that these do not contribute to the inversion formula.
As a result, we can reconstruct the full defect CFT data from the discontinuity of the bulk identity:
\begin{align}
 \Disc_\xb \Gm(x, \xb)
 = \Disc_\xb \left( \frac{\sqrt{x \xb}}{(1-x)(1-\xb)}\right)^\Dp
 = 2i \sin (\pi \Dp) \left( \frac{\sqrt{x}}{1-x}\right)^\Dp 
   \left( \frac{\sqrt{\xb}}{\xb-1}\right)^\Dp .
\end{align}
Plugging the discontinuity in the LIF \eqref{eq:inv-formula-even}, one can obtain the defect spectrum and the OPE coefficients.
This is worked out in detail in \cite{Lemos:2017vnx}, the main result is that the defect spectrum is given by $\Delta_{s,n} = \Dp + |s| + 2n$ with the following OPE coefficients:
\begin{align}
\label{eq:mft-defect}
 \mu_{s,n}^\GFF(\Dp, d) 
 = \frac{( \Dp + 1 - d/2 )_n (\Dp)_{2 n+|s|}}{n! (n+|s|)! (\Dp+n+|s|+1-d/2 )_n} \, .
\end{align}
For now we assume that the LIF converges down to $s = 0$, and we come back to the problem of convergence in section \ref{sec:altern-bc}.
We would like to use the defect data, which is analytic in $s$, to consider a monodromy defect in a bulk GFF. 
As pointed out around equation \eqref{eq:monodromy-defect-spins}, one obtains a monodromy defect by allowing the transverse spin to take non-integer values $s \in -v+\mathbb{Z}$.
Since we know the full defect CFT data, we can try to resum it and obtain the full correlation function:
\begin{align}
\begin{split}
 \Gm^\GFF_{\Dp,d,v}(x, \xb)
  = \sum_{n=0}^\infty \sum_{s \in \mathbb Z -v} 
 \mu_{s,n}^\GFF(\Dp, d) 
 \hat f_{\Dp + |s| + 2n, s}(x, \xb) \, .
\end{split}
\end{align}
As a consistency check, we note that the trivial case with no monodromy defect $v = 0$, resums to the bulk identity as one would expect:
\begin{align}
 \Gm^\GFF_{\Dp,d,v=0}(x, \xb)
 = \sum_{m=0}^\infty \sum_{s\in \mathbb Z} 
 \mu_{m,s}^\GFF(\Dp, d) 
 \hat f_{\Dp + s + 2m, s}(x, \xb)
 = \left( \frac{\sqrt{x \xb}}{(1-x)(1-\xb)}\right)^\Dp \, .
\end{align}
In the sections below, we consider three simple cases where the two-point correlator $\Gm(x,\xb)$ can also be obtained in closed form.

\subsubsection{Free theory monodromy defect}
\label{sec:free-monodromy-def}

The first simplification is to consider free bulk fields, which have conformal dimension $\Delta_\phi^\free = (d-2)/2$.
In this case only the leading transverse-twist trajectory $n=0$ contributes to the defect expansion, see \eqref{eq:mft-defect}.
Ideally we would like to find 
$ \Gm_{d,v}^\free(x, \xb) \equiv \Gm_{(d-2)/2, d,v}^\GFF(x, \xb)$ for general values of $d$ and $v$, but this turns out to be hard.\footnote{After this paper was submitted to the \texttt{arXiv}, we have been made aware by Y. Linke that there exists a closed form expression for $\Gm_{d,v}^\free(x, \xb)$ in terms of Appell $F_1$ functions. The precise formula can be provided by the authors upon request.}
Fortunately, for even spacetime dimension $d = 4, 6, \ldots$ the calculation simplifies dramatically and one can obtain closed form expressions.
For example, the $d = 4$ correlator is \cite{Giombi:2021uae}
\begin{align}
\begin{split}
\label{eq:free-4d-corr}
 \Gm^\free_{4,v}(x,\xb)
 & = \frac{\sqrt{x \xb}}{(1-x) (1-\xb)} 
     \frac{(1-\xb) x^v + (1-x) \xb^{1-v}}{1-x \xb} \, .
\end{split}
\end{align}
Similar expressions, though more lengthy, can be obtained for higher even values of $d$.
Keeping only the leading terms as $x \to 1$, the expressions simplify and it is possible to guess a formula for the correlator which is analytic in $d$
\begin{align}
\label{eq:free-lightcone-corr}
 \Gm_{d,v}^\free(x, \xb)
 = & \left( \frac{\sqrt{x \xb}}{(1-x) (1-\xb)} \right)^{\Delta^\free_\phi}
 \bigg(1 + \\ 
 & + C_{d,v}^\free \big( (1-x)(1-\xb) \big)^{\Delta^\free_\phi} 
 \bigg[ {}_2F_1
  \bigg( \! \begin{array}{*{20}{c}}
    {\Delta^\free_\phi, \Delta^\free_\phi+v } \\
    {d-1}
    \end{array};1-\xb \bigg) + O(1-x) \bigg] \bigg) \, , \nonumber
\end{align}
where we introduced the constants
\begin{align}
 C^\GFF_{\Dp,d,v}
 = -\frac{\Gamma (\Dp+1-v) \Gamma (\Dp+1+v)}{(\Dp+v) \Gamma (2 \Dp+1) \Gamma (v) \Gamma (1-v)} \, , \qquad
 C^\free_{d,v} = C^\GFF_{(d-2)/2,d,v} \, .
\end{align}
Even though \eqref{eq:free-lightcone-corr} has been obtained by non-rigorous means, it passes a number of non-trivial consistency checks.
It is correct for any even $d = 4,6,\ldots$, it is consistent with the result \cite{Liendo:2019jpu} for $v=1/2$ and general $d$, and it is consistent with the result \eqref{eq:mft-oeps} in $d=4-\veps$ dimensions.

The power of equation \eqref{eq:free-lightcone-corr} is that it captures all the bulk CFT data.
Indeed, since the bulk theory is free, the spectrum consists of double-twist operators $\Delta_{\ell,0} = 2 \Delta_\phi^\free + \ell$, namely
\begin{align}
\label{eq:free-bulk-exp}
 \Gm^\free_{d,v}(x, \xb)
 = \left( \frac{\sqrt{x \xb}}{(1-x) (1-\xb)} \right)^{(d-2)/2} \left( 
    1
    + \sum_{\ell=0}^\infty c^\free_\ell f_{\ell+d-2,\ell}(x, \xb)
 \right) \, .
\end{align}
Here we remind the reader that we use the shorthand notation $c_{\Om} = \lambda_{\phi\bar\phi\Om} a_\Om$.
Using the bulk blocks \eqref{eq:ligh-exp-bos-blocks} and comparing \eqref{eq:free-lightcone-corr}-\eqref{eq:free-bulk-exp} at leading order in $(1-x)$, one can obtain the bulk CFT data order by order in $(1-\xb)$.
For the first few coefficients we find
\begin{align}
\label{eq:first_cL_coeffs}
\begin{split}
 c_0^\free & = C^\free_{d,v}, \\
 c_1^\free & = \frac{(d-2) (2 v-1)}{4 (d-1)} C^\free_{d,v}, \\
\end{split}
\begin{split}
 c_2^\free & = \frac{(d-2) (v-1) v}{8 (d-1)} C^\free_{d,v}, \\
 c_3^\free & = \frac{(d-2) (d+2) (v-1) v (2 v-1)}{96 (d-1) (d+1)} C^\free_{d,v} \, .
\end{split}
\end{align}
The first three coefficients are in perfect agreement with the explicit calculation of \cite{Giombi:2021uae} up to a difference in normalization.\footnote{The value of $c_3$ also agrees with \texttt{v2} of \cite{Giombi:2021uae}.}
The main advantage of knowing the correlation function is that we can extract the bulk data for very high values of the spin $\ell$.
In doing this, we observed the CFT data satisfies a simple two-step recursion relation
\begin{align}
\label{eq:non-susy-rec}
  c_{\ell+2}^\free 
  = \frac{(2 v-1) (d+2 \ell)}{4 (\ell+2) (d+\ell)} c_{\ell+1}^\free
  + \frac{(\ell-1) (d+\ell-3) (d+2 \ell-2) (d+2 \ell)}
         {16 (\ell+2) (d+\ell) (d+2 \ell-3) (d+2 \ell-1)} c_\ell^\free \, ,
\end{align}
with the initial conditions as given in \eqref{eq:first_cL_coeffs}.\footnote{For $d=4$ we managed to obtain a closed-form expression by inverting the exact correlator \eqref{eq:free-4d-corr}:
\begin{align}
\begin{split}
 c_\ell^\free 
 \stackrel{d=4}{=} &
 \frac{\Gamma (\ell-1) \Gamma (\ell+1)^2 \sin ^2(\pi  v)}
       {2^{4 \ell+1} \pi  \Gamma (\ell+\frac{1}{2}) 
        \Gamma(\ell+\frac{3}{2}) \Gamma(\ell-v+1) \Gamma (\ell+v)}
 \Bigg[ \\
 & \quad \Gamma (2-v) \Gamma (\ell+v) \,
    {}_3F_2\left( {\begin{array}{*{20}{c}}
    {\ell+1, \ell+1, \ell-1 } \\
    {2(\ell+1), \ell-v+1}
    \end{array};1} \right)
   + (-1)^\ell \big( v \leftrightarrow 1-v \big) 
 \Bigg].
\end{split}
\end{align}
}

\subsubsection{Alternate boundary condition}
\label{sec:altern-bc}

The inversion formula predicts $\Delta_s = \Delta_\phi^\free + |s|$ for the free theory defect spectrum. 
However, as we pointed in section \ref{sec:inv-form}, this result only holds for spins $|s| > s_*$, where the threshold spin $s_*$ cannot be fixed from the bootstrap perspective.
In this subsection, we relax the assumption $s_* = 0$ for defects in free theories, which we show is related to continuing the correlator as $v \to v + n$ for $n \in \mathbb{Z}$.

In a free theory, the bulk equations of motion imply the defect spectrum is of the form
\begin{align}
 \Delta^\pm_s
 = \frac{d-2}{2} \pm |s| \, .
\end{align}
The positive modes $\Delta^+_s$ are given by the inversion formula, while the negative modes $\Delta^-_s$ can arise as low transverse-spin ambiguities for $|s| < s_*$.\footnote{In the setup of \cite{Giombi:2021uae}, the values $\Delta^\pm_s$ correspond to the two possible boundary conditions certain KK modes can have on the boundary of hyperbolic space $H^{d-1}$. We borrow the name of the section from this reference.}
The negative modes were studied in great detail in \cite{Lauria:2020emq,Behan:2020nsf} (see also \cite{Bianchi:2019sxz,Bianchi:2021snj}).
The outcome of these works is that if both $\Delta_s^+$ and $\Delta_s^-$ are present, the resulting defect is non-trivial.
Since we are interested on free defects, let us assume that for $s = -v$ we have a negative mode instead of a positive mode.
To obtain the correlator we substract the positive mode and add the negative one:
\begin{align}
\label{eq:def-altern-corr}
 \Gm^{\free,-}_{d,v}(x, \xb)
 = \Gm^{\free}_{d,v}(x, \xb)
 - \frac{\Gamma(\Delta_\phi^\free+v)}{\Gamma (\Delta_\phi^\free) \Gamma (1+v)}
   \hat f_{\Delta_{-v}^+, -v}(x, \xb)
 + \mu_{-v}^{\free,-}
   \hat f_{\Delta_{-v}^-, -v}(x, \xb) \, .
\end{align}
The OPE coefficient $\mu_{-v}^{\free,-}$ cannot be obtained with the inversion formula because this operator lies outside the range of convergence. 
Instead, we determine the OPE coefficient indirectly by demanding that $\Gm^{\free,-}_{d,v}(x, \xb)$ has a consistent bulk-channel expansion.
To achieve this, we expand the correlator to leading order in $1-x$ and order by order in $1-\xb$:
\begin{align}
\label{eq:raw-expansion}
\begin{split}
 \Gm^{\free,-}_{d,v}(x, \xb)
 = & \left( \frac{\sqrt{x \xb}}{(1-x) (1-\xb)} \right)^{(d-2)/2} \Bigg[  
    1 \\
  & + \big[(1-x)(1-\xb)\big]^{\frac{d-2}{2}} 
      \left( k_0 + k_1 (1-\xb) + k_2 (1-\xb)^2 + \ldots \right) \\
  & + \left(\frac{1-x}{1-\xb}\right)^{\frac{d-2}{2}} 
      \left( q_0 + q_1 (1-\xb) + q_2 (1-\xb)^2 +\ldots \right)
    + O\big((1-x)^{\frac{d}{2}} \big)
 \Bigg] \, .
\end{split}
\end{align}
The constants $k_i$ and $q_i$ can be determined to high order expanding \eqref{eq:def-altern-corr} using a computer algebra software. 
At the same time, because we are considering a free theory, we know the bulk spectrum consists of double-twist operators, so the block expansion takes the form:
\begin{align}
\label{eq:block-exp-altern}
 \Gm^{\free,-}_{d,v}(x, \xb)
 = \left( \frac{\sqrt{x \xb}}{(1-x) (1-\xb)} \right)^{(d-2)/2} \left( 
    1
    + \sum_{\ell=0}^\infty c^{\free,-}_\ell f_{\ell+d-2,\ell}(x, \xb)
 \right) \, .
\end{align}
Perhaps unexpectedly, these two expansions are inconsistent with each other, because the powers $q_i (1-\xb)^i$ in \eqref{eq:raw-expansion} cannot be reproduced from the blocks in \eqref{eq:block-exp-altern}.
The only way out is that the unknown OPE coefficient should takes the value
\begin{align*}
 \mu_{-v}^{\free,-}
 = \frac{\Gamma(\Delta_\phi^\free-v)}{\Gamma (\Delta_\phi^\free) \Gamma (1-v)} \, ,
\end{align*}
in which case $q_i = 0$ for $i \ge 0$, rendering the bulk expansion consistent.
This formula for $\mu_{-v}^{\free,-}$ is in perfect agreement with the explicit calculation of \cite{Giombi:2021uae}.

Now, the $x \to 1$ limit of the free correlator is given by \eqref{eq:free-lightcone-corr}, using hypergeometric identities one can combine \eqref{eq:free-lightcone-corr} with \eqref{eq:def-altern-corr} to obtain
\begin{align}
\label{eq:corr-altern-bc}
 \Gm_{d,v}^{\free,-}&(x, \xb)
 = \left( \frac{\sqrt{x \xb}}{(1-x) (1-\xb)} \right)^{\Delta^\free_\phi}
 \bigg(1 + \\ 
 & C_{v+1,d} \big( (1-x)(1-\xb) \big)^{\Delta^\free_\phi} 
 \bigg[ {}_2F_1
  \bigg( \! \begin{array}{*{20}{c}}
    {\Delta^\free_\phi, \Delta^\free_\phi+v+1 } \\
    {d-1}
    \end{array};1-\xb \bigg) + O(1-x) \bigg] \bigg) \, . \nonumber
\end{align}
Interestingly, this is just the original expression with the replacement $v \to v+1$.
Since \eqref{eq:corr-altern-bc} determines completely the bulk CFT data, and the bulk spectrum is independent of $v$, the full correlators satisfies the same relation:
\begin{align}
\begin{split}
 \Gm_{d,v}^{\free,-}(x, \xb) = \Gm_{d,v+1}^{\free}(x, \xb) \, .
\end{split}
\end{align}
As a result, the bulk OPE coefficients for alternate boundary conditions are obtained from \eqref{eq:first_cL_coeffs} by $v \to v+1$.
For spin $\ell =0,2$ we find perfect agreement with explicit calculation \cite{Giombi:2021uae}:
\begin{align}
\begin{split}
 c_0^{\free,-} = C_{v+1,d}, \qquad
 c_1^{\free,-} = \frac{(d-2) (2 v+1)}{4 (d-1)} C_{v+1,d} \, .
\end{split}
\end{align}
One can turn on more negative modes in a similar way. Note that in general these violate the defect unitarity bound, but this does not affect the discussion.
In particular, if we use negative modes for $s=-v,-v-1,\ldots,-v-n+1$, we find that the correlator is given by $\Gm^\free_{v+n,d}(x,\xb)$.
Similarly, if we turn on negative modes for $s=1-v,2-v,\ldots,n-v$ the correlator is given by $\Gm^\free_{v-n,d}(x,\xb)$.
More complicated choices of negative modes do not seem to generate such simple structure.

\subsubsection{GFF monodromy defect in \texorpdfstring{$d=4-\veps$}{d=4-eps}}
\label{sec:gff-monodromy-def}

In preparation for the analysis of the Wilson-Fisher fixed point, let us study GFF as a perturbation around free theory.
Consider a GFF scalar of dimension $\Dp = 1 - \delta_\phi \veps$ in $d=4-\veps$ dimensions. The defect data has been presented in equation \eqref{eq:mft-defect}.
In order to also extract bulk CFT data, it is necessary to resum the defect expansion.
The zeroth order result appears in \eqref{eq:free-4d-corr}, while here we carry out the resummation to leading order in $O(\veps)$.
For the leading transverse-twist family, there are contributions at $O(\veps)$ from the OPE coefficients, the defect blocks and the defect dimensions.
Furthermore, there are higher-twist families with $n > 0$ that only contribute with tree-level dimensions and OPE coefficients.
The complete $O(\veps)$ contribution is then:
\begin{align}
\label{eq:mft-oeps}
\begin{split}
 \Gm^{\GFF,O(\veps)}_{1 - \delta_\phi \veps,4-\veps,v}(x, \xb)
 & = \veps \sum_{n=0}^\infty \sum_{s \in -v + \mathbb{Z}}
 \partial_\veps \left( \mu_{s,n}^\GFF(1-\delta_\phi\veps, 4-\veps) \hat f_{1 + |s| - \delta_\phi \veps, s}(x, \xb) \right)_{\veps=0} \\
 & = -\delta_\phi \veps \frac{(x \xb)^{1/2}}{1-x\xb} \bigg[ 
   \frac{x^v}{1-x} 
      \left(
        \Phi(x, 1, v) + H_{v-1} + \log\left( \frac{\sqrt{x \xb}}{1-x\xb} \right) 
      \! \right) \\
 & \qquad \qquad +
     \frac{\xb x^{v}}{1-\xb} 
      \big( \Phi(x,1,v)-\Phi(x \xb,1,v) \big)
      + (x \leftrightarrow \xb, v \leftrightarrow 1-v) \bigg] \, . 
\end{split}
\end{align}
The result is written in terms of harmonic numbers $H_{n}$ and Hurwitz-Lerch zeta function $\Phi(x,1,v)$, which has nice properties reviewed in appendix \ref{sec:hurwitz-zeta}.
As a consistency check, for a free defect $\delta_\phi = 1/2$, the correlation function \eqref{eq:mft-oeps} at leading order in $x \to 1$ agrees with \eqref{eq:free-lightcone-corr} at leading order in $\veps$.
Let us also mention that there is a curious non-trivial cancellation of terms such that the final result is proportional to $\delta_\phi$.

We are now ready to expand in the bulk channel.
Once more, since the bulk theory is of the GFF type, the spectrum contains higher-twist families:
\begin{align}
 \Gm_{\Dp,d,v}^\GFF(x, \xb)
 = \left( \frac{\sqrt{x \xb}}{(1-x) (1-\xb)} \right)^{1-\delta_\phi\veps} \left(  
    1
    + \sum_{n=0}^\infty \sum_{\ell=0}^\infty c^\GFF_{\ell,n} f_{2\Dp+\ell+2n,\ell}(x, \xb)
 \right) \, .
\end{align}
As explained before, the bulk OPE coefficients can be extracted order by order in $(1-x)$,$(1-\xb)$ using the bulk blocks in the form \eqref{eq:ligh-exp-bos-blocks}.
Some of the low-lying coefficients are:
\begin{align}
\begin{split}
  & c^\GFF_{0,0} = C^\GFF_{\Dp,d,v} + O(\veps^2) \, , \\
  & c^\GFF_{1,0} = \frac{1}{18} C^\GFF_{\Dp,d,v} (2 v-1) (3 - \delta_\phi \veps)
                 + O(\veps^2) \, , \\
  & c^\GFF_{2,0} = \frac{1}{36} C^\GFF_{\Dp,d,v} v (v-1) (3 - \delta_\phi  \veps)
                 + O(\veps^2) \, , \\
  & c^\GFF_{0,1} = \frac{\veps}{96} (2 \delta_\phi -1) v (v-1) \left(v^2-v+4\right)
                 + O(\veps^2) \, .
\end{split}
\end{align}
The interested reader can find more OPE coefficients in the attached \mathematica notebook.

\subsection{Wilson-Fisher monodromy defect}

The last model we consider in this section is the $O(2N)$ Wilson-Fisher (WF) fixed point in $d=4-\veps$ dimensions.\footnote{The literature on the WF $O(N)$ model without defects is too vast to review here. However, let us mention the nice references \cite{Alday:2017zzv,Henriksson:2018myn}, which use analytic bootstrap techniques that inspired our work.}
Following \cite{Soderberg:2017oaa,Giombi:2021uae}, we impose a monodromy $v$ to the complex scalar $\phi = \phi_1 + i \phi_2$.
Besides the defect CFT data, we improve on existing results by computing the two-point function to order $O(\veps)$ and by extracting the bulk CFT data.
As already announced, we use the Lorentzian Inversion formula \eqref{eq:inv-formula}, which reconstructs defect CFT data from the discontinuity of the correlator $\Disc \Gm(x,\xb)$.
In perturbative CFTs, the discontinuity can be computed using information which is known from bulk physics at lower orders in perturbation theory.

As discussed in section \ref{sec:lif-applications}, only the bulk identity and double-twist operators with anomalous dimensions can contribute to the discontinuity.
For the Wilson-Fisher fixed point, this leads to dramatic simplifications that make it easy to bootstrap the correlator.
The key property is that the leading-twist trajectory has anomalous dimensions starting at order $O(\veps^2)$, i.e. $\Delta = 2 \Dp + \ell + O(\veps^2)$ for $\ell > 0$, and only the $\ell=0$ operator gets corrected at order $O(\veps)$:
\begin{align}
\label{eq:anom-dim-ppb}
 \Delta_{\phi\bar\phi}
 = 2 \Delta_\phi + \gamma_{\phi\bar\phi}^{(1)} \veps + O(\veps^2)
 = 2 \Delta_\phi + \frac{N+1}{N+4} \veps + O(\veps^2).
\end{align}
As a result, the discontinuity can be obtained to leading order $O(\veps)$ from a single bulk block 
\begin{align}
 \Disc_\xb \Gm(x,\xb) \big|_{O(\veps)}
 = \Disc_\xb \left(\frac{\sqrt{x \xb}}{(1-x)(1-\xb)} \right)^\Dp 
   \left(1 + f_{\Delta_{\phi\bar\phi},0}(x,\xb) \right) \, .
\end{align}
Notice that the bulk identity contribution has been studied separately in sections \ref{sec:free-monodromy-def} and \ref{sec:gff-monodromy-def}.
In particular, the external scalar has dimension $\Dp = (d-2)/2 + O(\veps^2)$ so this part of the correlator behaves as in free theory.

In what follows we neglect the identity contribution, and focus on the piece generated by the $\phi\bar\phi$ operator.
It has been described in section \ref{sec:lif-applications} how to compute the discontinuity.
In particular, combining equations \eqref{eq:compute-disc} and \eqref{eq:disc-cases}, expanding in $\veps$ and keeping only the $O(\veps)$ term we find
\begin{align}
 \Disc_x \Gm(x,\xb)
 = \Disc_\xb \Gm(x,\xb)
 = 2 \pi i \, \frac{\veps}{2} \frac{v(v-1)}{2} \frac{N+1}{N+4} \frac{(x \xb)^{1/2} \log(x \xb)}{1-x\xb} \, .
\end{align}
To obtain this discontinuity we also used the $d=4$ OPE coefficient $c_0^\free$ in \eqref{eq:first_cL_coeffs}, the anomalous dimension $\gamma_{\phi\bar\phi}^{(1)}$ in \eqref{eq:anom-dim-ppb}, and the bulk block $f_{2,0}$ from \eqref{eq:bulk-4d-blocks}.

Having derived the discontinuity, we are ready to extract the defect CFT data using the inversion formula.
Since the discontinuity is symmetric under $x \leftrightarrow \xb$, we can use the simpler formula \eqref{eq:inv-formula-even}, which applies to both positive and negative transverse spins.
Furthermore, since the discontinuity is $O(\veps)$, we can evaluate the LIF integration kernel exactly in $d=4$.
The resulting double integral is simple to do, giving
\begin{align}
\label{eq:wf-cft-data}
\begin{split}
 \mu(\Dh, s)
 & = \veps \frac{v (v-1)}{8} \frac{(N+1)}{(N+4)} 
     \int_0^1 dx \int_1^{1/x} d\xb \log (x \xb)
     x^{-\frac{\Dh-|s|+1}{2}} \xb^{-\frac{\Dh+|s|+1}{2}} \\
 & = - \veps \frac{v(v-1)}{2} \frac{(N+1)}{(N+4)} 
     \frac{1}{|s| \big( \Dh -|s|-1 \big)^2} \, .
\end{split}
\end{align}
It is well understood that in perturbative settings a double pole in $\Dh$ indicates defect anomalous dimensions, see \cite{Lemos:2017vnx} for details.
If one adds the contribution from the bulk identity to \eqref{eq:wf-cft-data}, then one concludes that the defect spectrum consists of a single family with CFT data
\begin{align}
\begin{split}
 \Dh_s 
 & = \frac{d-2}{2} + |s| + \veps \hat \gamma_s^{(1)} 
   + O(\veps^2) \, , \qquad
 \hat \gamma_s^{(1)} 
 = \frac{v(v-1)}{2} 
   \frac{(N+1)}{(N+4)}
   \frac{1}{|s|}\, , \\
 \mu_s
 & = \frac{(1-\veps/2)_{|s|}}{|s|!} + O(\veps^2)  \, ,
\end{split}
\end{align}
This is in perfect agreement with the literature \cite{Soderberg:2017oaa,Giombi:2021uae}.

Let us now extract the bulk OPE coefficients to order $O(\veps)$.
As for the free and GFF case, the first step is to resum the defect expansion.
The contribution of the bulk identity to the full correlator has been computed in equations \eqref{eq:free-4d-corr} and \eqref{eq:mft-oeps}, where one has to set $\delta_\phi = 1/2$ because $\phi$ behaves as a free field plus $O(\veps^2)$ corrections.
There is a contribution which is new for the Wilson-Fisher fixed point, which comes from the defect anomalous dimensions:
\begin{align}
\begin{split}
 \Gm_{\text{WF}}(x, \xb)
 & = \veps \sum_{s \in -v + \mathbb{Z}}
 b_{|s|}^2 \hat \gamma_{|s|}^{(1)} \partial_{\Dh} \hat f_{\Dh, s}(x, \xb)\big|_{\Dh=|s|+1} \\
 & = \veps \frac{v(v-1)}{4} \frac{(N+1)}{(N+4)}
     \frac{(x \xb)^{1/2} \log(x \xb)}{1-x\xb} \Big[ 
    x^v \Phi(x, 1, v) 
    + \xb^{1-v} \Phi(\xb, 1, 1-v) 
 \Big] \, .
\end{split}
\end{align}
For $v = 1/2$ and $N=1/2$, this reproduces the Ising $\mathbb Z_2$ monodromy defect result \cite{Liendo:2019jpu}.

We have obtained the full two-point correlation function to $O(\veps)$, so it is now an easy exercise to extract the bulk OPE coefficients.
Besides the twist-two family there is also a twist-four family:\footnote{The absence of higher-twist families at this order was suggested in \cite{Liendo:2012hy} for $\ell = 0$, and then proven in \cite{Alday:2017zzv}.}
\begin{align}
\begin{split}
 \Gm^\free_{d,v}(x, \xb)
 + \Gm_{\text{WF}}(x, \xb)
 & = \left( \frac{\sqrt{x \xb}}{(1-x) (1-\xb)} \right)^{\Dp} \Bigg( 
    1
    + c_{0,0} f_{d-2+\veps\gamma_{\phi\bar\phi}^{(1)},0}(x, \xb) \\
 &  \qquad \qquad
    + \sum_{\ell=1}^\infty c_{\ell,0} f_{\ell+d-2,\ell}(x, \xb)
    + \sum_{\ell=0}^\infty c_{\ell,1} f_{\ell+4,\ell}(x, \xb)
 \Bigg) \, .
\end{split}
\end{align}
The OPE coefficients of the leading-twist trajectory take a particularly simple form after normalizing by the free piece
\begin{align}
\begin{split}
 c_{0,0} 
 & = c_{0}^\free \left( 1 + \frac{\veps}{2} \frac{(N+1) }{(N+4)} (H_{v-1} + H_{-v})
     + O(\veps^2) \right) \, , \\
 c_{1,0} 
 & = c_{1}^\free \left( 1 + \frac{3\veps}{2} \frac{(N+1)}{(N+4)} + O(\veps^2) \right) \, , \\
 c_{2,0} 
 & = c_{2}^\free \left( 1 + \frac{\veps}{6} \frac{(N+1)}{(N+4)} 
        \frac{(3 v-2) (3 v-1)}{v(v-1)} + O(\veps^2) \right) \, , \\
 c_{3,0} 
 & = c_{3}^\free \left( 1 + \frac{\veps}{6} \frac{(N+1)}{(N+4)} 
 \frac{\left(10 v^2-10 v+3\right)}{v(v-1)} + O(\veps^2) \right) \, .
\end{split}
\end{align}
On the other hand, the subleading-twist trajectory has the following CFT data:
\begin{align}
\begin{split}
 c_{0,1} 
 & = \frac{\veps}{16} \frac{(N+1)}{(N+4)} v^2 (v-1)^2 + O(\veps^2) \, , \\
 c_{1,1} 
 & = \frac{\veps}{144} \frac{(N+1)}{(N+4)} v^2 (v-1)^2 (2 v-1) + O(\veps^2) \, , \\
 c_{2,1} 
 & = \frac{\veps}{1920} \frac{(N+1)}{(N+4)} v^2 (v-1)^2 \left(5 v^2-5 v+2\right) + O(\veps^2) \, .
\end{split}
\end{align}
All our results are in perfect agreement with the Ising $\mathbb Z_2$ monodromy defect \cite{Liendo:2019jpu}.
The interested reader can find the bulk OPE coefficients for higher values of $\ell$ in the attached \mathematica notebook.

Before concluding, let us remind the reader that the bulk OPE coefficients are defined as $c_\Om = \lambda_{\phi\bar\phi\Om} a_\Om$, where $a_\Om$ is proportional to the one-point function of $\Om$.
Therefore, one can obtain the one-point functions of leading-twist operators as $a_{\ell,0} = c_{\ell,0} / \lambda_{\phi\bar\phi \Om_{\ell,0}}$, where the three-point OPE coefficient $\lambda_{\phi\bar\phi \Om_{\ell,0}}$ is well know at order $O(\veps)$ \cite{Dey:2016mcs,Henriksson:2018myn}.
Unfortunately, the twist-four trajectory contains nearly-degenerate operators, so our OPE coefficient has to be interpreted as a sum over these operators
\begin{align}
 c_{\ell,1} 
 = \langle\!\langle a_{\ell,1} \lambda_{\phi\bar\phi \Om_{\ell,1}} \rangle\!\rangle
 \equiv \sum_{\text{deg. ops. } \Om^{(n)}} 
 a^{(n)}_{\ell,1} \lambda^{(n)}_{\phi\bar\phi \Om_{\ell,1}} \, .
\end{align}
In this case, the best we can do is to extract an average density of one-point OPE coefficients defined as $\langle\!\langle a_{\ell,1} \rangle\!\rangle = \langle\!\langle a_{\ell,1} \lambda_{\phi\bar\phi \Om_{\ell,1}} \rangle\!\rangle / \langle\!\langle \lambda^2_{\phi\bar\phi \Om_{\ell,1}} \rangle\!\rangle^{1/2}$, where once again the average over three-point OPE coefficients $\langle\!\langle \lambda^2_{\phi\bar\phi \Om_{\ell,1}} \rangle\!\rangle$ is known  \cite{Henriksson:2018myn}.

\section{Wess-Zumino: Bulk theory}
\label{sec:wess-zumino-bulk}

Superconformal field theories (SCFTs) in non-integer dimensions were studied in \cite{Bobev:2015jxa,Bobev:2015vsa}, where the numerical bootstrap gave evidence that the Wess-Zumino model \eqref{eq:wz-lagrangian} is perhaps the simplest SCFT preserving four supercharges.
In this section we study the Wess-Zumino model in $d = 4-\veps$ dimensions (without defects) using the analytic bootstrap,
and the results will be needed for the study of defects in section \ref{sec:wess-zumino-defect}. We work to leading order in $O(\veps)$, but the same methods also work at higher orders in $\veps$, a subject that we plan to study in future work.
For the reader that is mostly interested on the final results, we present a self-contained summary of the CFT data in section \ref{sec:wz-summary}.

\subsection{Generalities}

Let us briefly review some generalities of SCFT in non-integer dimensions, more details can be found in \cite{Bobev:2015jxa}.\footnote{A different type of superconformal theories in non-integer dimensions also appear in the context of Parisi-Sourlas supersymmetry \cite{Kaviraj:2019tbg,Kaviraj:2020pwv}}
The conformal part of the algebra is generated by the usual operators $D$, $P_i$, $K_i$ and $M_{ij}$ with $i = 1, \ldots, d$. 
There are exactly four Poincaré supercharges $Q^+_\a$ and $Q^-_\ad$ and four conformal supercharges $S^{\ad+}$ and $S^{\a-}$, where the indices take two values $\a,\ad=1,2$ regardless of the spacetime dimension. The supercharges obey the usual supersymmetry algebra
\begin{align}
 \{ Q^+_\a, Q^-_\ad \} = \Sigma^i_{\a\ad} P_i\,, \qquad
 \{ S^{\ad+}, S^{\a-} \} = \bar \Sigma_i^{\ad\a} P_i \,.
\end{align}
There is also a generator $R$ of $U(1)_R$ symmetry, under which $Q^+_\a$ and $Q^-_\ad$ have charge $+1$ and $-1$ respectively.
The monodromy defects in section \ref{sec:wess-zumino-defect} will be naturally obtained by twisting this $U(1)_R$ symmetry.

In what follows, we focus our attention on chiral-primary operators $\phi$ and their complex conjugates $\bar \phi$.
These operators are killed by supercharges of the same chirality, and the superconformal algebra fixes their conformal dimension in terms of their $R$-charge:
\begin{align}
\label{eq:chiral-primary}
   \left[ Q^+_\a, \phi(0) \right] 
 = \left[ Q^-_\ad, \bar\phi(0) \right]
 = 0 \quad \Rightarrow \quad
   \Delta_{\phi} 
 = \Delta_{\bar\phi} 
 = \frac{d-1}{2} R_\phi
 = - \frac{d-1}{2} R_{\bar\phi}\,.
\end{align}

In order to bootstrap the Wess-Zumino model without defects, we consider four-point functions  of $\phi$ and $\bar \phi$.
If we focus on the $s$-channel expansion, there are three inequivalent orderings of the external operators:
\begin{align}
\begin{split}
 & \langle \phi(x_1) \phi(x_2) \bar \phi(x_3) \bar \phi(x_4) \rangle
 = \frac{\Fm(z, \zb)}{(x_{12}^2 x_{34}^2)^\Dp} 
 = \frac{1}{(x_{12}^2 x_{34}^2)^\Dp}
   \sum_{\Delta,\ell \text{ even}} a_{\Delta,\ell} \, g_{\Delta,\ell} (z, \zb)\, , \\
 & \langle \phi(x_1) \bar \phi(x_2) \phi(x_3) \bar \phi(x_4) \rangle
 = \frac{\Gm(z, \zb)}{(x_{12}^2 x_{34}^2)^\Dp} 
 = \frac{1}{(x_{12}^2 x_{34}^2)^\Dp} 
   \sum_{\Delta,\ell} b_{\Delta,\ell} \,
   G_{\Delta,\ell}(z, \zb)\, , \\
 & \langle \bar \phi(x_1) \phi(x_2) \phi(x_3) \bar \phi(x_4) \rangle
 = \frac{\tilde \Gm(z, \zb)}{(x_{12}^2 x_{34}^2)^\Dp}
 = \frac{1}{(x_{12}^2 x_{34}^2)^\Dp}
   \sum_{\Delta,\ell} (-1)^\ell b_{\Delta,\ell} \, \tilde G_{\Delta,\ell}(z, \zb)\, .
\end{split}
\end{align}
In the above formula $a_{\Delta,\ell}$ and $b_{\Delta,\ell}$ are shorthand notation for three-point OPE coefficients squared.
The three orderings above are related to each other by simple crossing relations:
\begin{align}
\label{eq:susy-crossing}
\begin{split}
 & \Gm(z, \zb)
 = \left( \frac{z \zb}{(1-z)(1-\zb)} \right)^{\Dp} \Gm(1-z, 1-\zb) \, , \\
 & \Fm(z, \zb)
 = \left( \frac{z \zb}{(1-z)(1-\zb)} \right)^{\Dp} \tilde \Gm(1-z, 1-\zb) \, .
\end{split}
\end{align}
The functions $\Gm(z,\zb)$ and $\tilde \Gm(z,\zb)$ capture the same CFT data in their $s$-channel expansion, since they are related by $1 \leftrightarrow 2$.

The constraints of supersymmetry are accounted for by expanding the correlation function in terms of superconformal blocks \cite{Bobev:2015jxa}.
It can be shown that in the $\phi \times \phi$ OPE superconformal blocks reduce to regular non-supersymmetric blocks $g_{\Delta,\ell}$.
On the other hand, the superblocks $G_{\Delta,\ell}$ are non-trivial for the $\phi \times \bar\phi$ OPE.
Interestingly, in any dimension the superblocks take the simple form of non-supersymmetric blocks for unequal external operators with a suitable prefactor:
\begin{align}
\label{eq:superblocks-shift}
\begin{split}
 G_{\Delta,\ell}(z, \zb)
 = (z \zb)^{-1/2} g^{1,1}_{\Delta+1,\ell} \, , \qquad
 \tilde G_{\Delta,\ell}(z, \zb)
 = (z \zb)^{-1/2} g^{1,-1}_{\Delta+1,\ell} \, .
\end{split}
\end{align}
Superconformal blocks capture the contributions to the OPE of all exchanged operators that belong to the same supermultiplet, which means they should decompose as finite sums of non-supersymmetric blocks with relative coefficients fixed by susy.
This is indeed the case
\begin{align}
\label{eq:superblocks-explicit}
\begin{split}
 & G_{\Delta,\ell}(z, \zb)
 = g_{\Delta,\ell} 
 + a_1 \, g_{\Delta+1,\ell+1} 
 + a_2 \, g_{\Delta+1,\ell-1} 
 + a_3 \, g_{\Delta+2,\ell} \, , \\
 & \tilde G_{\Delta,\ell}(z, \zb)
 = g_{\Delta,\ell} 
 - a_1 \, g_{\Delta+1,\ell+1} 
 - a_2 \, g_{\Delta+1,\ell-1} 
 + a_3 \, g_{\Delta+2,\ell} \, , \\
\end{split}
\end{align}
where the explicit coefficients are
\begin{align}
\label{eq:coeffs-blocks}
\begin{split}
 & a_1 = \frac{(\Delta+\ell)}{4(\Delta+\ell+1)} \, , \\
 & a_2 = \frac{\ell(\ell+d-3)(\Delta-\ell-d+2)}
              {(2\ell+d-4)(2\ell+d-2)(\Delta-\ell-d+3)} \, , \\
 & a_3 = \frac{\Delta (\Delta-d+3)(\Delta+\ell)(\Delta-\ell-d+2)}
              {4(2\Delta-d+4)(2\Delta-d+2)(\Delta+\ell+1)(\Delta-\ell-d+3)} \, .
\end{split}
\end{align}

\subsubsection{Comments on degenerate operators}

There is an important difference between the Wilson-Fisher fixed point studied in \cite{Alday:2017zzv,Henriksson:2018myn} and Wess-Zumino model studied here, which is the existence of nearly-degenerate operators in the leading-twist family $[\phi\bar\phi]_{\ell,0}$.
Indeed, from the Lagrangian \eqref{eq:wz-lagrangian} it is clear that we can construct two leading-twist operators for $\ell > 0$:
\begin{align}
\label{eq:two-ops}
\begin{split}
 \Om_{\ell,0}^{(1)} 
 \sim k_{11} \phi \partial_{\mu_1} \ldots \partial_{\mu_\ell} \bar \phi 
 +    k_{12} \psi \partial_{\mu_1} \ldots \partial_{\mu_{\ell-1}} \sigma_{\mu_\ell} \psi^\dag 
 \, , \\
 \Om_{\ell,0}^{(2)} 
 \sim k_{21} \phi \partial_{\mu_1} \ldots \partial_{\mu_\ell} \bar \phi
    + k_{22} \psi \partial_{\mu_1} \ldots \partial_{\mu_{\ell-1}} \sigma_{\mu_\ell} \psi^\dag \, .
\end{split}
\end{align}
The coefficients $k_{nm}$ are fixed demanding that the operators $\Om^{(n)}_{\ell,0}$ are conformal-primary operators, with well-defined scaling dimensions in the interacting theory, and orthonormal with respect to two-point functions.
Near the free theory, when the anomalous dimensions $\gamma_{\ell,0}^{(n)}$ are small, the expansion of the four-point function in conformal blocks has to be interpreted as a sum over nearly-degenerate operators
\begin{align}
 \Gm(z,\zb)
 \sim p_{0,0} g_{2\Dp,0} 
 + \sum_{\ell=1}^\infty \left(
    \langle\!\langle p_{\ell,0} \rangle\!\rangle g_{2\Dp+\ell,\ell}
    + \langle\!\langle p_{\ell,0} \gamma_{\ell,0} \rangle\!\rangle \partial_\Delta g_{2\Dp+\ell,\ell}
    + \ldots \right)
 + \ldots \, .
\end{align}
In the previous equation higher-twist operators are neglected, and the expansion coefficients are sums over the two operators in \eqref{eq:two-ops}:
\begin{align}
 \langle\!\langle p_{\ell,0} \rangle\!\rangle 
 = \sum_{n=1,2} p_{\ell,0}^{(n)} \, , \qquad
 \langle\!\langle p_{\ell,0} \gamma_{\ell,0} \rangle\!\rangle 
 = \sum_{n=1,2} p_{\ell,0}^{(n)} \gamma_{\ell,0}^{(n)} \, , \qquad
 \text{etc.}
\end{align}
Although we have focused on the leading-twist trajectory for clarity, similar complications also occur with higher-twist trajectories.

For a general CFT, it would be quite challenging to solve this mixing problem using bootstrap techniques.
Fortunately, the supersymmetry of the Wess-Zumino model allows for a simple resolution.
The main observation is that, in terms of supersymmetry representations, one of the combinations in \eqref{eq:two-ops} is a superprimary operator, while the other is a superdescendant operator.
This can be checked with the superconformal blocks \eqref{eq:superblocks-explicit}, noticing that for each superprimary operator with dimension $(\Delta, \ell)$, there is a superdescendant operator with equal twist and one more unit in spin $(\Delta+1, \ell+1)$.
For example, in free theory the first operator in the $\phi \times \bar\phi$ OPE is the superprimary $\phi\bar\phi$ with $(\Delta,\ell) = (2\Dp,0)$. 
Then, the descendant $(\phi\bar\phi)_{\text{desc}}$ with quantum numbers $(2\Dp+1, 1)$ will be degenerate with a superprimary $\Om^{\text{prim}}_{1,0}$ with the same quantum numbers.
Continuing in this way, the descendant of $\Om^{\text{prim}}_{1,0}$ will be degenerate with the superprimary $\Om^{\text{prim}}_{2,0}$, and so on and so forth.

The moral of the story is that for the Wess-Zumino model, the degeneracies in the leading-twist family can be understood as arising from the supersymmetry of the model.
Therefore, by using a superconformal block expansion
\begin{align}
 \Gm(z, \zb)
 = \sum_{\Delta,\ell} b_{\Delta,\ell} \, G_{\Delta,\ell}(z, \zb)\, , 
\end{align}
it is guaranteed that all degeneracies in the leading-twist family are taken into account.
In other words, we have argued that the OPE coefficients $b_{\ell,0}$ capture the contributions of individual superprimary operators.
If one is interested in the contribution of a certain superdescendant, it is then sufficient to use the superconformal blocks \eqref{eq:superblocks-explicit} to relate it to the superprimary.
On the other hand, we expect the OPE coefficients of higher-twist families $b_{\ell,n\ge1}$ to be sums over nearly-degenerate operators.

\subsection{Inversion formula}
The next tool we need are inversion formulas, which reconstruct the CFT data from certain discontinuities of correlators \cite{Caron-Huot:2017vep}.
The main object of interest are functions that encode dimensions as poles and OPE coefficients as residues:
\begin{align}
 a_{\Delta,\ell} = - \Res_{\Delta'=\Delta} a(\Delta', \ell) \, , \qquad
 b_{\Delta,\ell} = - \Res_{\Delta'=\Delta} b(\Delta', \ell).
\end{align}
Let us start with the inversion formula that reconstructs $a(\Delta, \ell)$.
Since the $\phi \times \phi$ OPE uses non-supersymmetric blocks, we can use the inversion formula originally derived by Caron-Huot \cite{Caron-Huot:2017vep}:
\begin{align}
\label{eq:inv-form-nonsus}
\begin{split}
 & a(\Delta, \ell) 
 = \frac{1 + (-1)^\ell}{4} \kappa^{0,0}_{\Delta+\ell}
   \int_0^1 \int_0^1 \frac{dz d\zb}{(z \zb)^d}
   \left| z-\zb \right|^{d-2} 
   g_{\ell+d-1, \Delta+1-d}(z, \zb) \dDisc[\Fm(z, \zb)] \, . \\
\end{split}
\end{align}
The double discontinuity is defined in the usual way
\begin{align}
  & \dDisc[\Fm(z, \zb)]
 = \Fm(z, \zb) 
 - \frac{1}{2} \Fm(z, \zb^\circlearrowleft) 
 - \frac{1}{2} \Fm(z, \zb^\circlearrowright) \, ,
\end{align}
where the analytic continuation is performed around the branch point $\zb = 1$ in the directions indicated by the arrows.
The overall constant has the following value
\begin{align}
\begin{split}
 & \kappa_{2\hb}^{2r,2s}
 = \frac{\Gamma(\hb+r)\Gamma(\hb-r)\Gamma(\hb+s)\Gamma(\hb-s)}
        {2\pi^2 \Gamma(2\hb-1) \Gamma(2\hb)} \, .
\end{split}
\end{align}

Similarly, there exists an inversion formula that reconstructs $b(\Delta,\ell)$.
In order to obtain it, note that superconformal blocks are non-supersymmetric blocks with shifted arguments \eqref{eq:superblocks-shift}.
Using the inversion formula for completely general external operators \cite{Caron-Huot:2017vep,Simmons-Duffin:2017nub}, after some manipulations we find
\begin{align}
\label{eq:inv-form-sus}
\begin{split}
 b(\Delta, \ell)
 = \frac{\kappa^{1,1}_{\Delta+\ell+1}}{4}
    \int_0^1 \int_0^1 \frac{dz d\zb}{(z \zb)^d} 
&   |z - \zb|^{d-2} 
    \bigg(
    g^{1,1}_{\ell+d-1, \Delta-d+2}(z, \zb) 
    \dDisc\big[(z \zb)^{1/2} \Gm(z,\zb)\big] \\
&
 + (-1)^{\ell+1}
    g^{-1,1}_{\ell+d-1, \Delta-d+2}(z, \zb) 
    \dDisc \big[ (z \zb)^{1/2} \tilde \Gm(z,\zb) \big]
 \bigg) \, .
\end{split}
\end{align}
A simple way to see that the $t$- and $u$-channel contributions must be different is to note that the superconformal blocks used in the expansion of $\Gm(z,\zb)$ and $\tilde \Gm(z,\zb)$ are different \eqref{eq:superblocks-shift}.

In practice, it is convenient to expand the integrand of the inversion formulas in the limit $z \to 0$ and integrate term by term.
In the limit $z \to 0$ the correlator has an expansion of the following form
\begin{align}
\label{eq:Fsing}
\begin{split}
 \Fm(z, \zb)
& = \sum_{n=0}^\infty \sum_{p=0}^\infty \,
    z^{\Dp+n} \log^p \! z \, \Fm_{n,p}(\zb) \, ,
\end{split}
\end{align}
and similarly for $\Gm(z,\zb)$ and $\tilde \Gm(z,\zb)$.
The inversion formula integration kernels can also be expanded in the limit $z \to 0$:
\begin{align}
 \frac{1}{z} \left( \frac{\zb - z}{z \zb} \right)^{d-2} 
 g^{r,s}_{\ell+d-1,\Delta+1-d}(z, \zb)
 = z^{-(\Delta-\ell)/2}
   \sum_{m=0}^\infty \sum_{j=-m}^m 
   \Cm^{r,s}_{m,j}(\Delta,\ell) z^m k^{r,s}_{\Delta+\ell + 2j}(\zb) \, .
\end{align}
Similarly to equation \eqref{eq:ligh-exp-bos-blocks}, the coefficients in this expansion can be fixed recursively using the four-point Casimir equation.
This type of expansion has been described in detail in the appendix of \cite{Caron-Huot:2017vep,Liu:2020tpf}. 
After expanding the inversion formula as above, the only non-trivial integrals left to do are of the form
\begin{align}
\label{eq:inv-integral}
\begin{split}
 \INV[g(\zb)](\beta) 
 & = \int_0^1 \frac{d\zb}{\zb^2} k_{\beta}(\zb)
     \dDisc \big[ g(\zb) \big] \, , \\
 \SINV^\pm[g(\zb)](\beta) 
 & = \int_0^1 \frac{d\zb}{\zb^{3/2}} k^{\pm1,1}_{\beta+1}(\zb)
     \dDisc \! \big[ g(\zb) \big] \, .
\end{split}
\end{align}
Finally, the last integral in $z$ is elementary and produces poles in $\Delta$.

Collecting the ingredients together, we have obtained new versions of the Lorentzian inversion formula.
For $a(\Delta,\ell)$ we find
\begin{align}
\label{eq:non-susy-inv}
\begin{split}
 & a(\Delta, \ell)
 = - \sum_{n,p = 0}^\infty 
     \frac{S_{n, p}(\Delta, \ell)}{(\Delta - \Delta_\phi - \ell - 2n)^{p+1}} \, , \\
 & S_{n, p}(\Delta, \ell)
 = \big( 1 + (-1)^\ell \big) 2^p p! \, \kappa_{\Delta+\ell}^{0,0} 
   \sum_{m=0}^n \sum_{k=-m}^m
   \Cm^{0,0}_{m,k}(\Delta, \ell)
   \INV[\Fm_{n-m, p}(\zb)](\Delta+\ell+ 2k) \, .
\end{split}
\end{align}
Similarly, one obtains $b(\Delta,\ell)$ using the following formula:
\begin{align}
\label{eq:susy-inv}
\begin{split}
 & b(\Delta, \ell)
 = - \sum_{n,p = 0}^\infty 
     \frac{S_{n, p}(\Delta, \ell)}{(\Delta - \Delta_\phi - \ell - 2n)^{p+1}} \, , \\
 & S_{n, p}(\Delta, \ell)
 = 2^p p! \, \kappa_{\Delta+\ell+1}^{1,1} 
   \sum_{m=0}^n \sum_{k=-m}^m \Bigg[
   \Cm^{1,1}_{m,k}(\Delta+1, \ell)
   \SINV^+[\Gm_{n-m, p}(\zb)](\Delta+\ell+ 2k) \\
 & \hspace{10.5em}
 + (-1)^{\ell+1} 
   \Cm^{-1,1}_{m,k}(\Delta+1, \ell)
   \SINV^-[\tilde \Gm_{n-m, p}(\zb)](\Delta+\ell+ 2k) 
 \Bigg] \, .
\end{split}
\end{align}
These new formulas are simpler to use in perturbative settings, such as the ones we consider in this paper.

\subsection{Generalized free field theory}
\label{sec:gff-bulk}

As a first application of the inversion technology, let us consider generalized free field theory (GFF).
In order to extract the CFT data $a_{\Delta,\ell}$ in the $\phi \times \phi$ OPE we have to use the GFF correlation function
\begin{align}
\begin{split}
 \Fm(z, \zb) & = (z \zb)^\Dp + \left( \frac{z \zb}{(1-z)(1-\zb)} \right)^\Dp \, .
\end{split}
\end{align}
The first term is regular around $\zb = 1$ so it is killed by the discontinuity and it does not contribute to the inversion formula.
Expanding in $z \to 0$ and using the definition \eqref{eq:Fsing} we find
\begin{align}
\label{eq:sing-part-gff}
\begin{split}
 \Fm(z, \zb)|_{\text{singular}}
 = \left( \frac{z \zb}{(1-z)(1-\zb)} \right)^{\Dp} \quad \Rightarrow \quad
 \Fm_{n,p}(\zb) 
 = \delta_{p,0} \frac{(\Dp)_n}{n!} \left( \frac{\zb}{1-\zb}\right)^\Dp.
\end{split}
\end{align}
The next step is to compute the integral \eqref{eq:inv-integral}.
A useful trick is to use the Euler representation of the hypergeometric function, and swap the order of integration.
The result is \cite{Caron-Huot:2017vep}:
\begin{align}
\begin{split}
 \INV \left[ \left( \frac{\zb}{1-\zb}\right)^p \, \right](\beta)
 & = 2 \pi^2 \frac{\Gamma (\beta)}{\Gamma (\beta/2)^2}
     \frac{\Gamma (\beta/2 + p - 1)}{\Gamma (p)^2 \Gamma (\beta/2 - p +1)}.
\end{split}
\end{align}
All the ingredients can be combined using equation \eqref{eq:non-susy-inv} to obtain the dimensions and OPE coefficients for low values of $n$.
We find the family of operators $[\phi\phi]_{\ell,n}$ with dimension $\Delta_{\ell,n} = 2\Dp + \ell + 2n$ and their OPE coefficients agree with the results of \cite{Fitzpatrick:2011dm}:
\begin{align}
\label{eq:gff-abulk}
\begin{split}
 a^{\GFF}_{\ell,n}(\Dp, d) 
 & = \frac{2 \left(\Dp+1-d/2\right)_n^2 (\Dp)_{\ell+n}^2}
   {\ell! n! \left(\ell + d/2\right)_n (2 \Dp+n+1-d)_n (2 \Dp+\ell+2 n-1)_\ell } \\
 & \hspace{16em}
   \times \frac{1}{\left(2 \Dp+\ell+n - d/2\right)_n}.
\end{split}
\end{align}

A similar calculation allows one to obtain the OPE coefficients in the $\phi \times \bar\phi$ OPE.
Now the relevant GFF correlation functions are
\begin{align}
\begin{split}
        \Gm(z, \zb) = 1 + \left( \frac{z \zb}{(1-z)(1-\zb)} \right)^{\Dp} \, , \qquad
 \tilde \Gm(z, \zb) = 1 + (z \zb)^\Dp \, .
\end{split}
\end{align}
Clearly $\Gm(z,\zb)$ has the same singular part as $\Fm(z,\zb)$, see equation \eqref{eq:sing-part-gff}, while $\tilde \Gm(z,\zb)$ is regular around $\zb = 1$ and does not contribute to the LIF.
Using similar techniques as before one obtains the following integral
\begin{align}
 \SINV^+ \left[ \left( \frac{\zb}{1-\zb}\right)^p \, \right](\beta)
 & = 2 \pi ^2 
   \frac{\Gamma (\beta +1)}{\Gamma (\beta/2 +1)^2}
   \frac{\Gamma (\beta/2 + p)}{\Gamma (p)^2 \Gamma (\beta/2 -p+1)} \, .
\end{align}
Once again, using \eqref{eq:susy-inv} one can obtain the first few OPE coefficients $b_{\ell,n}$ of the operators $[\phi \bar \phi]_{\ell,n}$.
They are in perfect agreement with the values reported in \cite{Bobev:2015jxa}
\begin{align}
\label{eq:gff-bbulk}
\begin{split}
 b^\GFF_{\ell,n}(\Dp, d)
 & = \frac{\left(\Dp+1-d/2\right)_n^2 (\Dp)_{\ell+n}^2}
   {\ell! n! \left(\ell + d/2\right)_n (2 \Dp+n+2-d)_n (2 \Dp+\ell+2 n)_\ell} \\
 & \hspace{15em}
   \times \frac{1}{\left(2 \Dp+\ell+n+1 - d/2\right)_n} \, .
\end{split}
\end{align}

\subsection{Wess-Zumino model}

We are now ready to solve the Wess-Zumino model at leading order in $\veps = 4-d$.
There is a well-known Lagrangian formulation for this model \eqref{eq:wz-lagrangian},  which consists of a single chiral superfield $\Phi$ interacting with cubic superpotential $\Wm \sim \Phi^3$.
In this section we follow a bootstrap approach similar to \cite{Alday:2017zzv}, but it is useful to keep in mind the Lagrangian \eqref{eq:wz-lagrangian}.
At the end, we check that our results are in perfect agreement with the literature.

\subsubsection{A family of solutions to crossing}

At order $O(\veps^0)$ the theory consists of a free chiral multiplet in $d=4$.
The spectrum and OPE coefficients can be obtained from the previous section by setting $\Dp = 1$.
In particular, formulas \eqref{eq:gff-abulk} and \eqref{eq:gff-bbulk} imply that only the leading double-twist families $n=0$ contribute. 
When we turn on interactions for small $\veps$, the dimension of the external chiral gets corrected $\Delta_\phi = 1 - \delta_\phi \veps + O(\veps^2)$.
Furthermore, the operators in the two OPEs $\phi\times\phi$ and $\phi\times\bar\phi$ can also get corrected, and new families of operators could appear in the OPEs.

Let us start studying the $\phi\times\phi$ CFT data at the next order $O(\veps)$.
The LIF \eqref{eq:inv-form-nonsus} reconstructs the CFT data from the discontinuity of $\Fm(z,\zb)$.
Using the crossing equation \eqref{eq:susy-crossing}, the discontinuity can be computed in terms of the $\phi \times \bar \phi$ CFT data.
There is one contribution from the bulk identity, which is considered in section \ref{sec:gff-bulk}, and a contribution from anomalous dimensions.
The corrections from anomalous dimensions are of order $O(\veps^2)$ and can be neglected.
Since the inversion formula is not expected to converge for low values of $\ell$, we should also include a term $\Hm(z, \zb)$ with finite support in spin:
\begin{align}
\begin{split}
 \Fm(z, \zb) & 
 = (z \zb)^\Dp 
 + \left( \frac{z \zb}{(1-z)(1-\zb)} \right)^\Dp 
 + \veps \Hm(z, \zb) \, .
\end{split}
\end{align}
Solutions to crossing with finite support in spin were studied in \cite{Alday:2016jfr}, and it was found that around $d=4$ there is one such solution that takes the form
\begin{align}
 \Hm(z, \zb) 
 = k \big( 1 - \partial_\Delta \big) g^{d=4}_{\Delta,0}(z, \zb) \big|_{\Delta=2} \, .
\end{align}
For now the constant $k$ should be treated as an unknown, but later its value will be fixed.
This correlator has the following decomposition in conformal blocks
\begin{align}
 \Fm(z, \zb)
 = 
   \left(a_{0,0}^{(0)} + \veps a_{0,0}^{(1)} \right) 
   g_{2\Dp+\veps\gamma,0}
  + \sum_{\substack{\ell=2 \\ \ell \text{ even}} }^\infty 
   a_{\ell, 0}
   g_{2\Dp+\ell,\ell}
 + \sum_{\substack{\ell=0 \\ \ell \text{ even}} }^\infty  
   a_{\ell, 1} g_{2\Dp+2+\ell,\ell} \, .
\end{align}
Notice there is a new family of twist-four operators with tree-level OPE coefficients.
To the order we are working, we have $a_{\ell, n} = a_{\ell, n}^\GFF(\Dp,d)$.
The only exception is the $\ell=n=0$ case, when the $[\phi\phi]_{0,0}$ operator has the following CFT data:
\begin{align}
 a_{0,0} = 2 + \veps k \, , \qquad
 \gamma = - \frac k2 \, .
\end{align}

Let us now turn to the CFT data in the $\phi \times \bar\phi$ OPE.
The inversion formula \eqref{eq:inv-form-sus} has a $t$-channel contribution and a $u$-channel contribution.
As before, one uses the crossing equation \eqref{eq:susy-crossing} and the OPE expansion to see which terms contribute.
The $t$-channel contribution consists of the identity, which has been studied in section  \ref{sec:gff-bulk}, and anomalous dimensions that contribute at order $O(\veps^2)$.
An unfamiliar feature of the supersymmetric inversion formula \eqref{eq:inv-form-sus} is that the $u$-channel contribution produces $O(\veps)$ corrections to the CFT data.
Using crossing, the part of $\tilde \Gm(z,\zb)$ proportional to $\log(1-\zb)$ is given by the $[\phi\phi]_{0,0}$ operator we just studied:\footnote{Here $g_{\Delta,\ell}(z,\zb) = (z \zb)^{(\Delta-\ell)/2} \tilde g_{\Delta,\ell}(z, \zb)$ is defined analogously to \eqref{eq:red-blocks}.}
\begin{align}
\label{eq:gtild-sing}
\begin{split}
 \tilde \Gm(z,\zb) \big|_{\log(1-\zb)}
 & = \frac{\veps}{2} a_{0,0} \gamma (z \zb)^\Dp \log(1-\zb) 
     \tilde g_{2,0}(1-z, 1-\zb) \\
 & = -\frac{\veps}{2} k (z \zb)^\Dp \log(1-\zb) \frac{\log z - \log \zb}{z-\zb} \, .
\end{split}
\end{align}
From this result, it is clear that the only inversions integrals that one needs to do are:
\begin{align}
\label{eq:uchann-inv}
\begin{split}
 & \SINV^- \! \big[ \zb^{-n} \log(1-\zb) \big](\beta)
 = \frac{2 \pi ^2 \Gamma (\beta+1)}{\Gamma (\beta/2+1)^2} \, , \\
 & \SINV^- \! \big[ \zb^{-n} \log(1-\zb) \log \zb \big](\beta)
 = 0 \, .
\end{split}
\end{align}
In order to obtain these inversions, we expand the integrand in powers of $(1-\zb)/\zb$, integrate term by term, and in the end resum an asymptotic expansion in powers of $1/\beta$.
This procedure has been explained in detail in \cite{Alday:2017zzv,Alday:2019clp}, where the reader can find further details.
The ingredients \eqref{eq:gtild-sing}-\eqref{eq:uchann-inv} can be combined using \eqref{eq:susy-inv} to find $b(\Delta,\ell)$.
We find that to this order in $\veps$, the $\phi \times \bar\phi$ OPE consists only of the leading-twist family
\begin{align}
 \Gm(z, \zb)
 = 1 
 + \sum_{\ell=0}^\infty 
   b_{\ell, 0}
   G_{2\Dp+\ell+\veps\gamma_\ell,\ell}
 + O(\veps^2) \, ,
\end{align}
where the CFT data can be readily obtained using the inversion formula
\begin{align}
\label{eq:bulk-wz-cft-data-k}
 \gamma_\ell = k \frac{(-1)^{\ell+1}}{\ell+1} \, , \qquad
 b_{\ell,0}
 = b_{\ell,0}^\GFF(\Dp,d) \left( 1 
 + k (-1)^{\ell+1} \frac{ \left(H_\ell - H_{2\ell+1}\right)}{(\ell+1)} \veps
 + O(\veps^2) \right) \, .
\end{align}
An important observation is that this result makes sense even for spin $\ell = 0$.
Furthermore, we expect the Lorentzian inversion formula to have better convergence properties in supersymmetric theories \cite{Lemos:2021azv}.
Thus, we make the plausible assumption that \eqref{eq:bulk-wz-cft-data-k} is valid for all $\ell \ge 0$.

\subsubsection{Fixing the coefficients}

We have found a two-parameter family of solutions to crossing which depend on $k$ and $\delta_\phi$, let us now try to fix these coefficients from basic physical requirements.
The first condition is that the stress tensor is conserved.
The stress tensor belongs to a short multiplet with a superprimary of dimensions $\Delta=d-1$ and spin $\ell=1$, as can be seen from the form of the superconformal block:
\begin{align}
\label{eq:stress-tensor-block}
 G_{d-1,1}
 = g_{d-1,1}
 + \frac{d}{4(d+1)} g_{d,2}\,.
\end{align}
As a result, conservation of the stress tensor requires that the operator $[\phi\bar\phi]_{1,0}$ has dimension $d-1$.
This relates $\delta_\phi$ and $k$ as follows
\begin{align}
 2 \Delta_\phi + 1 + \veps \gamma_1 = d-1 
 \qquad \Rightarrow \qquad
 \delta_\phi = \frac{k+2}{4} \, .
\end{align}
On the other hand, the identification of the operator $[\phi\phi]_{0,0}$ allows to fix the remaining free parameter.
As it was discussed in \cite{Bobev:2015jxa}, this operator can be identified with a chiral-primary operator $\phi^2$, in which case:
\begin{align}
 [\phi\phi]_{0,0} = \phi^2 
 \quad \Rightarrow \quad
 2 \Delta_\phi + \veps \gamma_0 = 2 \Dp 
 \quad \Rightarrow \quad
 k = 0, \;
 \delta_\phi = \frac{1}{2} \, .
\end{align}
We conclude that if $[\phi\phi]_{0,0} = \phi^2$ the theory is free in $d=4-\veps$ dimensions.

A second possibility discussed in \cite{Bobev:2015jxa} is that $[\phi\phi]_{0,0}$ is a level-two descendant of $\bar \phi$:
\begin{align}
 [\phi\phi]_{0,0} = (Q^+)^2 \bar \phi 
 \quad \Rightarrow \quad
 2 \Delta_\phi + \veps \gamma_0 = \Dp + 1
 \quad \Rightarrow \quad
 k = - \frac{2}{3}, \;
 \delta_\phi = \frac{1}{3} \, .
\end{align}
This leads to a non-vanishing $k$, so we have found a non-trivial supersymmetric CFT in $d=4-\veps$ dimensions.
In the following section we provide evidence that this CFT is indeed the Wess-Zumino model.

\subsubsection{Summary and discussion}
\label{sec:wz-summary}

Let us summarize our results on the Wess-Zumino model at order $O(\veps)$.
The first result of our bootstrap analysis is the dimension of the external chiral field:
\begin{align}
\label{eq:dimPhi}
 \Dp = \frac{d-1}{3} \, .
\end{align}
This is actually a well-known result.
Recall that the Wess-Zumino model has a cubic superpotential $\Wm \sim \Phi^3$, which must have $R$-charge $R_\Wm = 2$ at the fixed point.
As a result, the chiral-primary field $\phi(x)$ must have charge $R_\phi = 2/3$, or equivalently $\Dp = (d-1)/3$, which means that \eqref{eq:dimPhi} is in fact an exact result to all orders in $\veps$.

The $\phi \times \phi$ OPE consists of double-twist operators $[\phi \phi]_{\ell,n}$, which are of the schematic form $\phi \Box^n \partial_{\mu_1} \ldots \partial_{\mu_\ell} \phi$.
The two families $n=0,1$ contribute at order $O(\veps)$, with CFT data given by the GFF results in section \ref{sec:gff-bulk}.
The only exception is the $[\phi \phi]_{0,0}$ operator, which has the following CFT data:
\begin{align}
 a_{0,0} = 2 - \frac{2}{3} \veps + O(\veps^2) \, , \qquad
 \Delta_{0,0} = 2 \Dp + \frac{\veps }{3} + O(\veps^2) \, .
\end{align}
The first observation is that $\Delta_{0,0} \ne 2 \Dp$ so we cannot interpret $[\phi\phi]_{0,0}$ as a chiral-primary operator $\phi^2$.
This is consistent because the Wess-Zumino model has a chiral ring relation $\phi^2 = 0$ due to the cubic superpotential.
Instead, the correct interpretation is $[\phi\phi]_{0,0} = (Q^+)^2 \bar\phi$, which agrees with our results since $\Delta_{0,0} = \Dp + 1$ and the $R$-charge is conserved.
The presence of such an operator is consistent with the OPE selection rules \cite{Bobev:2015jxa}, and it was also suggested by the numerical bootstrap results of \cite{Bobev:2015vsa}.
Thus, we expect the relation $\Delta_{0,0} = \Dp + 1$ to hold to all orders in $\veps$.

The $\phi \times \bar \phi$ OPE contains superconformal primaries and superconformal descendants, and their precise contribution can be obtained from the superconformal blocks \eqref{eq:superblocks-explicit}.
We expect superprimaries of the schematic form $\Om_\ell = \phi \partial_{\mu_1} \ldots \partial_{\mu_\ell} \bar \phi + \psi \partial_{\mu_1} \ldots \partial_{\mu_{\ell-1}} \sigma^{\mu_\ell}\bar \psi$, where the precise relative coefficients should be fixed by demanding $S^{\pm}\Om_\ell = 0$.
From our bootstrap analysis we found the following CFT data:
\begin{align}
\label{eq:bulk-wz-cft-data}
\begin{split}
 b_{\ell}
 & = b_{\ell,0}^\GFF(\Dp,d) \left( 
   1 
 + (-1)^\ell \frac{2 \left(H_\ell - H_{2\ell+1}\right)}{3 (\ell+1)} \veps
 + \Om(\veps^2)
 \right) \, , \\
 \Delta_{\ell} 
 & = 2 \Dp + \ell + \frac{2}{3} \frac{(-1)^{\ell}}{\ell+1} \veps
  + O(\veps^2) \, .
\end{split}
\end{align}
It is natural to identify the $\ell = 0$ operator with $\phi \bar \phi$, which has dimension $\Delta_{\phi\bar\phi} = 2 + O(\veps^2)$ \cite{Fei:2016sgs}, in perfect agreement with our results.
Finally, using \eqref{eq:stress-tensor-block} one can relate the OPE coefficient $b_1$ to the central charge\footnote{We define the central charge as in \cite{Poland:2018epd}, such that the stress-tensor contribution to the OPE is of the form:
$\langle \phi \bar \phi \phi \bar \phi \rangle \supset \frac14 (\frac{d}{d-1})^2 \frac{\Delta_\phi^2}{C_T} g_{d,2} \, $.
}
\begin{align}
 C_T 
 = \frac{d(d+1)}{(d-1)^2} \frac{\Delta_\phi^2}{b_1}
 = \frac{20}{3}-\frac{17 \veps }{9} + O(\veps^2) \, .
\end{align}
Once again this is in perfect agreement with the literature \cite{Fei:2016sgs}, up to a difference in normalization.

\section{Wess-Zumino: Monodromy defects}
\label{sec:wess-zumino-defect}

In this section we generalize the analysis of section \ref{sec:wf-monodromy} to superconformal theories with four Poincare supercharges. We study half-BPS monodromy defects that preserve two Poincare supercharges and focus on two-point functions of chiral operators.
We start the section with general results valid for monodromy defects in arbitrary superconformal theories, and then move on to the specific case of a monodromy defect for the Wess-Zumino model studied in section \ref{sec:wess-zumino-bulk}.

\subsection{Superconformal blocks}

Let us start by calculating the relevant superconformal blocks. We use techniques originally developed for bulk four-point functions \cite{Bobev:2015jxa,Bobev:2017jhk} and later applied to superconformal boundaries \cite{Gimenez-Grau:2020jvf}.
Here we only give an outline the calculation, the interested reader can find further details in the aforementioned references. We stress again that this section applies to general half-BPS codimension-two defects, which need not be monodromy defects.

\subsubsection{Defect superconformal algebra}

As in section \ref{sec:wf-monodromy}, we chose our codimension-two defect to sit at $x^1 = x^2 = 0$. 
The subalgebra of conformal transformations that preserve the defect is generated by $D$, $P_a$, $K_a$ and $M_{ab}$, where $a,b = 3, \ldots, d$ are indices parallel to the defect.
Since translation symmetry is partly broken, at most half of the original supercharges can be preserved by the defect.
Following the conventions of section \ref{sec:wess-zumino-bulk}, we choose the preserved supercharges to be:
\begin{align}
\label{eq:defect-supercharges}
 \Qm_1 = Q^+_1 \,, \qquad
 \Qm_2 = Q^-_1 \,, \qquad
 \Sm_1 = S^{1+} \,, \qquad
 \Sm_2 = S^{1-} \,.
\end{align}
Using the following Clifford algebra representation $\Sigma^i_{\a\ad} = (\bar \Sigma_i^{\ad\a})^* = (\sigma_1, \sigma_2, \sigma_3, i \mathds{1})$, it is possible to check in $d = 3$ and $d=4$ that the supercharges generate a subalgebra of the full superconformal algebra. 
For non-integer dimensions $3 \le d \le 4$ this construction is less rigorous, however we will obtain perfectly consistent results.
The anticommutators of the supercharges generate translations and special conformal transformations parallel to the defect:
\begin{align}
 \{ \Qm_A, \Qm_B \} = \widehat \Sigma_{AB}^a P_a\,, \qquad
 \{ \Sm_A, \Sm_B \} = \widehat \Sigma_{AB}^a K_a\,, \qquad 
 a = 3, \ldots, d, \quad
 A, B = 1,2 \, .
\end{align}
Similarly, by considering anticommutators of the form $\{ \Qm, \Sm \}$, we observe that the defect does not preserve $R$--symmetry or transverse rotations independently, but only a particular linear combination of them:\footnote{The full subalgebra for $d=3$ can be found in \cite{Agmon:2020pde} in conventions slightly different to ours.}
\begin{align}
\label{eq:twist-trans-spin}
 \Mm = M_{12} + \frac{d-1}{2} R \,.
\end{align}
With these conventions in mind, we proceed to obtain the superconformal blocks.

\subsubsection{Defect channel}

Let us start with the defect OPE $\phi(x) \sim \sum \widehat \Om(\vec y)$.
In this channel only one operator per defect supermultiplet contributes to the OPE, and as a result, the defect superconformal blocks $\hat F_{\Dh,s}(x,\xb)$ reduce to bosonic blocks $\hat f_{\Dh,s}(x,\xb)$.
In our conventions $\Dh,s$ label the conformal primary exchanged in the OPE, and not the superprimary in the corresponding multiplet.

We justify the above claim following an argument from \cite{Poland:2010wg}.
Since the chirality condition \eqref{eq:chiral-primary} is preserved by the defect supercharges \eqref{eq:defect-supercharges}, it turns out that $[\Qm_1, \phi(x)] = [\Sm_1, \phi(x)] = 0$.
Inserting these relations in the OPE implies $[\Qm_1, \widehat \Om(\vec y)] = [\Sm_1, \widehat \Om(\vec y)] = 0$.
However, only one operator in each defect supermultiplet can satisfy both of these conditions, hence superblocks in this channel are just standard bosonic blocks.

\subsubsection{Bulk channel}
\label{sec:bulk-4eps}

In the bulk channel, up to four conformal primaries in each supermultiplet can contribute to the OPE.
Their contributions are organized in superconformal blocks which we now calculate.

Following \cite{Dolan:2003hv,Fitzpatrick:2014oza}, we characterize superconformal blocks as solutions to the supersymmetric Casimir equation.
The superconformal Casimir can be split naturally into a non-supersymmetric and a supersymmetric piece: $C_{\text{full}} = C_{\text{bos}} + C_{\text{susy}}$.
The first contribution $\frac12 C_{\text{bos}}$ leads to the differential operator in equation \eqref{eq:cas-eq}.
The second contribution is due to supersymmetry:
\begin{align}
\label{eq:supercas}
\begin{split}
 C_{\text{susy}} 
 & =
 - \frac{d-1}{2} R^2
 + \frac{1}{2} [S^{\ad+}, Q^-_\ad]
 + \frac{1}{2} [S^{\a-}, Q^+_\a]\,.
\end{split}
\end{align}
Following \cite{Bobev:2015jxa}, our goal is to massage \eqref{eq:supercas} into a differential operator that can be added to \eqref{eq:cas-eq}.
Using the commutation relations, the chirality properties of $\phi$ and $\bar \phi$, and equation (51) from~\cite{Bobev:2015jxa} we find:
\begin{align}
\begin{split}
 \left[ C_{\text{susy}}, \phi(x_1) \bar \phi(x_2) \right] |0\rangle
 & = i x_{12}^\mu \bar \Sigma_\mu^{\ad\a}
     \left[ Q^-_\ad,      \phi_1(x_1) \right] 
     \left[ Q^+_\a , \bar \phi_2(x_2) \right] |0\rangle
   + 4 \Dp \phi(x_1) \bar \phi(x_2) |0\rangle \,.
\end{split}
\end{align}
Using superconformal Ward identities as in \cite{Bobev:2015jxa,Gimenez-Grau:2020jvf} to rewrite the $Q$-dependent part as a differential operator we get
\begin{align}
\label{eq:blk-contrib-PPb}
 \frac12 C_{\text{susy}} \langle \phi_1(x_1) \bar \phi_2(x_2) \rangle
 \to
 -\big[(1-x) \partial_x
 + \xb (1-\xb) \partial_\xb \big] F_{\Delta,\ell}(x,\xb) \,.
\end{align}
Combining the bosonic equation \eqref{eq:cas-eq}, the supersymmetric one \eqref{eq:blk-contrib-PPb}, and using the appropriate supersymmetric eigenvalue $c_2 = \Delta  (\Delta - d + 2) + \ell (\ell+d-2)$, we obtain a differential equation for the superconformal block $F_{\Delta,\ell}(x,\xb)$.
In $d=4$ the solution with correct boundary conditions takes a simple form:
\begin{align}
\begin{split}
 F_{\Delta,\ell}(x, \xb)
 = \frac{\sqrt{(1-x)(1-\xb)}}{1 - x\xb} 
 & \Big( 
     k_{\Delta-\ell-1}^{1,-1}(1-x) k_{\Delta+\ell+1}^{1,1}(1-\xb) \\
 & + (-1)^\ell k_{\Delta+\ell+1}^{1,-1}(1-x) k_{\Delta-\ell-1}^{1,1}(1-\xb) 
 \Big) \,.
\end{split}
\end{align}
For general $d$, we use an expansion of the form
\begin{align}
\label{eq:lightcone-series-superblocks}
 F_{\Delta,\ell}(x, \xb)
 = \sum_{n=0}^\infty \sum_{j=-n}^n B_{n,j}(\Delta,\ell) 
   (1-x)^{(\Delta-\ell)/2 + n} (1-\xb)^{-1/2} k_{\Delta+\ell+1+2j}^{1,1}(1-\xb) \, ,
\end{align}
and we fix the coefficients using the supercasimir equation.
The procedure is easy to implement using a computer algebra system.
For the first few coefficients we find:
\begin{align}
B_{0,0}(\Delta,\ell) = 1 \, , \quad
B_{1,-1}(\Delta,\ell) = \frac{(2-d) \ell}{d+2 \ell-4} \, , \quad
B_{1,1}(\Delta,\ell) = \frac{(2-d) \Delta  (\Delta +\ell) (\Delta +\ell+2)}{16 (2 \Delta +4-d) (\Delta +\ell+1)^2} \, .
\end{align}
Finally, let us mention that the superconformal block has a decomposition into a sum of four bosonic blocks:
\begin{align}
 F_{\Delta,\ell}(x, \xb)
 = f_{\Delta,\ell}(x, \xb) 
 + a_1 \, f_{\Delta+1,\ell+1}(x, \xb) 
 - a_2 \, f_{\Delta+1,\ell-1}(x, \xb) 
 - a_3 \, f_{\Delta+2,\ell}(x, \xb)  \, .
\end{align}
The coefficients can be found in \eqref{eq:coeffs-blocks}.
The fact that the coefficients are the same as the four-point blocks of chiral operators might seem surprising at first.
Actually, with the identification \eqref{eq:map-xz} the defect bulk blocks $F_{\Delta,\ell}(x,\xb)$ are the analytic continuation of the four-point blocks $\tilde G_{\Delta,\ell}(z,\zb)$ \cite{Gimenez-Grau:2020jvf}.
What we have found is that the close connection between codimension-two defects and four-point functions also holds at the superconformal level.

\subsection{Free and GFF half-BPS monodromy defect}
\label{sec:gff-superdef}

Armed with the superconformal blocks, we can now bootstrap superconformal monodromy defects.
In this section we focus on defects in (generalized) free theories, while we leave the more interesting defect in the Wess-Zumino model for the next section.
Fortunately, we can recycle many results from the non-supersymmetric case studied in section \ref{sec:wf-monodromy}.

Let us start with the case of a free bulk theory preserving four supercharges.
Since $\phi(x)$ is a free-field, its correlation function $\Gm^\free_{d,v}(x, \xb)$ is independent of the rest of the field content of the theory, so it is given by the non-supersymmetric formulas \eqref{eq:free-4d-corr}-\eqref{eq:free-lightcone-corr}.
Moreover, the defect superblocks reduce to non-supersymmetric blocks, so the defect CFT data is given by \eqref{eq:mft-defect}.
The story is more interesting in the bulk channel, because now in order to obtain the CFT data one must use superconformal blocks:
\begin{align}
 \Gm^\free_{d,v}(x, \xb)
 = \left( \frac{\sqrt{x \xb}}{(1-x) (1-\xb)} \right)^{(d-2)/2} \left(  
    1
    + \sum_{\ell=0}^\infty d_\ell^\free F_{d-2+\ell,\ell}(x, \xb)
 \right) \, .
\end{align}
Once again, we use the shorthand notation $d_\Om = \lambda_{\phi\bar\phi\Om}a_\Om$. 
Since the bulk theory is free, only the leading-twist family contributes. 
Using the series representation \eqref{eq:lightcone-series-superblocks} for the superblocks, we can extract the CFT data order by order in $(1-x)$ and  $(1-\xb)$.
For the first few coefficients we find:
\begin{align}
\begin{split}
 d_0^\free = C^\free_{d,v}, \quad
 d_1^\free = \frac{(d-2) (v-1)}{2 (d-1)} C^\free_{d,v}, \quad
 d_2^\free = \frac{(d-2) (v-1) (d v-d+v)}{8 (d-1) (d+1)} C^\free_{d,v} \, .
\end{split}
\end{align}
Similarly to section \ref{sec:wf-monodromy}, the coefficients satisfy a two-step recursion relation which can be used to efficiently go to high values of $\ell$:\footnote{Once again, in the $d = 4$ case it is possible to obtain a closed analytic formula:
\begin{align}
\begin{split}
 d_\ell^\free
 \stackrel{d=4}{=} 
   \frac{\Gamma (\ell) \Gamma (\ell+1) \Gamma (\ell+2) \sin ^2(\pi  v)}
        {2^{4 \ell+3}\pi  \Gamma(\ell+3/2)^2} 
 \Bigg[
    \frac{\Gamma (2-v)}{\Gamma (\ell-v+2)}
    {}_3F_2\left( {\begin{array}{*{20}{c}}
    {\ell, \ell+1, \ell+2 } \\
    {2\ell+3, \ell-v+2}
    \end{array};1} \right) \\
    + (-1)^{\ell+1} \frac{\Gamma (v)}{\Gamma (\ell+v)}
    {}_3F_2\left( {\begin{array}{*{20}{c}}
    {\ell, \ell+1, \ell+1 } \\
    {2\ell+3, \ell+v}
    \end{array};1} \right)
 \Bigg] \, .
\end{split}
\end{align}}
\begin{align}
\begin{split}
 d^\free_{\ell +2}
 & = \frac{(d+2 \ell ) \left(d^2 (v-1)+d (4 v-3) \ell +d+(4 v-3) \ell ^2-v\right)}{2 (\ell +2) (d+\ell ) (d+2 \ell -1) (d+2 \ell +1)} d^\free_{\ell+1} \\
 & \quad +\frac{\ell  (d+\ell -2) (d+2 \ell -2) (d+2 \ell )}{16 (\ell +2) (d+\ell ) (d+2 \ell -1)^2} d^\free_{\ell} \, .
\end{split}
\end{align}

The next simplest example is a monodromy defect in a bulk GFF theory.
As in the free case, the full correlator $\Gm^\GFF_{\Dp,d,v}(x, \xb)$ is the same as in the non-supersymmetric theory, and the defect CFT data is given by \eqref{eq:mft-defect}.
For the bulk data we can use \eqref{eq:mft-oeps}, which is the leading order correlator in $\veps = 4-d$ around the free value $\Dp = 1 - \delta_\phi \veps$. Expanding in bulk blocks
\begin{align}
 \Gm^\GFF_{\Dp,d,v}(x, \xb)
 = \left( \frac{\sqrt{x \xb}}{(1-x) (1-\xb)} \right)^{\Dp} \left(  
    1
    + \sum_{n=0}^\infty \sum_{\ell=0}^\infty d_{\ell,n}^\GFF F_{2\Dp+\ell+2n,\ell}(x, \xb)
 \right) \, ,
\end{align}
it is relatively straightforward to extract CFT data up to high values of $\ell$ and $n$ using the expansion \eqref{eq:lightcone-series-superblocks}.
Some of the low-lying coefficients are
\begin{align}
\label{eq:bulk-gff-sus-opes}
\begin{split}
 d_{0,0}^\GFF & = C_{\Dp,d,v} + O(\veps^2) \, , \\
 d_{1,0}^\GFF & = \frac{1}{9} (v-1) (3 - \delta_\phi  \veps) C_{\Dp,d,v} + O(\veps^2) \, , \\
 d_{0,1}^\GFF & = \frac{\veps}{96} (2 \delta_\phi -1) v (v-1) (v-2) (v-3) + O(\veps^2) \, , \\
\end{split}
\end{align}
while we give more coefficients in the attached \mathematica notebook.

\subsection{Wess-Zumino model}

Finally, we proceed to bootstrap the two-point function of chiral operators in the Wess-Zumino model to order $O(\veps)$ in the $\veps$--expansion.
The derivation requires knowledge of the bulk theory derived in section \ref{sec:wess-zumino-bulk} and the inversion formula derived in section \ref{sec:inv-form}.
Although the calculations for the Wess-Zumino model are similar in spirit to the Wilson-Fisher fixed point, in practice they are more challenging and require extra technology which we develop in the appendix.

Let us remind the reader that the Wess-Zumino model is a theory of a single chiral superfield with cubic superpotential $\Wm \sim \Phi^3$.
At the fixed point, the chiral-primary field $\phi(x)$ must have charge $R_\phi = 2/3$, or equivalently $\Dp = (d-1)/3$.
Since the external dimension differs from free theory at order $O(\veps)$, there is a GFF contribution with $\delta_\phi = 1/3$, which has been discussed in section \ref{sec:gff-superdef}.

Furthermore, as discussed in section \ref{sec:wess-zumino-bulk}, the bulk OPE contains double-twist operators $[\phi\bar\phi]_{\ell,n}$.
Importantly, the leading-twist operators $n=0$ have OPE coefficients of order $O(1)$ and anomalous dimensions $\gamma_{\ell}$ of order $O(\veps)$, see \eqref{eq:bulk-wz-cft-data}.
As a result, the entire leading-twist family contributes to $\Disc \Gm(x,\xb)$.
Indeed, the part of the correlator with non-vanishing discontinuity is 
\begin{align}
\label{eq:sing-part-wz}
\begin{split}
 \Gm(x,\xb) |_{\text{singular}}
 & = \frac{\veps}{2} (x \xb)^{\Dp/2} \log \! \big[ (1-x)(1-\xb) \big]
   \sum_{\ell=0}^\infty d^\free_{\ell} \gamma_{\ell} \tilde F_{2\Dp+\ell, \ell}(x, \xb) \, \\
 & = - \frac{\veps}{3} v(v-1) (x \xb)^{1/2} \log \! \big[ (1-x)(1-\xb) \big]
 \frac{h \! \left(\frac{\xb-1}{\xb}\right) - h(1-x)}{1 - x \xb} \, ,
\end{split}
\end{align}
where we introduced $h(z) = z \, _3F_2(1,1,v+1;2,3;z)$.
From here it is in principle straightforward to extract the defect CFT data using the bulk-to-defect Lorentzian inversion formula.
However, for the sake of clarity, we defer the details to appendix \ref{sec:app-inversion-wz}.
Below we present the defect CFT data, which contains contributions from the bulk identity (GFF) and from \eqref{eq:sing-part-wz}.

\paragraph{Leading transverse-twist family:}

The first family are defect operators of transverse twist approximately one. 
Since these operators are present in the free theory, their conformal dimensions can get corrected at this order in perturbation theory:
\begin{align}
\begin{split}
\label{eq:dim-lead-trans-twist}
 \Dh_{s,0} & = \frac{d-1}{3} + |s| + \veps \hat \gamma_{s}^{(1)} + O(\veps^2) \, , \qquad
 \hat \gamma_{s}^{(1)}
 = \begin{cases}
    0 & \text{for } s>0 \, , \\
    \frac{2 (v-1)}{3 |s|} & \text{for } s<0 \, .
   \end{cases}
\end{split}
\end{align}
Furthermore, their OPE coefficients also get corrected as follows:
\begin{align}
\begin{split}
 & \mu_{s>0,0}
 = 1 
 + \frac{-(2 |s| + 1 - v)H_{|s|} + (|s|+1-v) H_{|s|+1-v} - (1-v) H_{1-v}}{3 |s|} \veps 
 + O(\veps^2)  \, , \\
 & \mu_{s<0,0}
 = 1 
 + \frac{-(2|s| + v - 1)H_{|s|} + (|s|+v-1) H_{|s|+v-1} - (v-1) H_{v-1}}{3 |s|}  \veps 
 + O(\veps^2)  \, .
\end{split}
\end{align}
An important feature of the CFT data is that it is not symmetric under $s \leftrightarrow -s$.
Even though this seems surprising at first, it follows because $\phi(x)$ is a complex field, complex conjugation relates positive transverse-spin modes from $\phi(x)$ with the negative modes from $\bar \phi(x)$.
From a technical point of view, this asymmetry is due to \eqref{eq:sing-part-wz} not being symmetric under $x \leftrightarrow \xb$.
In particular, one would observe a similar phenomena in the $O(N)$ Wilson-Fisher fixed point starting at order $O(\veps^2)$ and $N>1$.

\paragraph{Subleading transverse-twist families:}

The next families of defect operators have transverse twist $2n + 1$.
At this order in perturbation theory, only the tree-level dimensions contribute
\begin{align}
\begin{split}
 \Dh_{s,n} & = 1 + |s| + 2n  + O(\veps) \quad  \; \text{for} \quad n \ge 1 \, .
\end{split}
\end{align}
Notice that these families receive contributions both from the bulk identity and from \eqref{eq:sing-part-wz}, and as a result, the defect OPE coefficients differ from the GFF values:
\begin{align}
\label{eq:wz-def-norm}
\begin{split}
 \mu_{s>0,n}
 = \frac{|s| + 2 (1-v)}{6 n (|s| + n)} \veps + O(\veps^2) \, , \qquad
 \mu_{s<0,n}
 =\frac{|s| + 2 (v-1)}{6 n (|s| + n)}  \veps + O(\veps^2) \, .
\end{split}
\end{align}

\paragraph{Fractional transverse-twist families:}

Perhaps surprisingly, there is another family of defect operators with non-integer transverse twist.
Indeed, their tree-level conformal dimensions are
\begin{align}
\begin{split}
 \Dh_{s>0,n}^{\text{fr}} = 1 + |s| + 2(n + 1 - v) \, , \quad 
 \Dh_{s<0,n}^{\text{fr}} & = 1 + |s| + 2(n + v - 1) \, ,
 \qquad \text{for} \quad n \ge 1 \, .
\end{split}
\end{align}
Notice that this family is generated exclusively from the bulk leading-twist family \eqref{eq:sing-part-wz}.
Once more, the tree-level OPE coefficients take a rather simple form:
\begin{align}
\label{eq:wz-def-weird}
\begin{split}
 & \mu_{s>0,n}^{\text{fr}}
 = \frac{n}{3 (n+1-v) (|s| + n + 1-v)}\, , \quad
 \mu_{s<0,n}^{\text{fr}}
 =\frac{n}{3 (n+v-1) (|s| + n + v-1)} \, .
\end{split}
\end{align}

Having reviewed the structure of the defect CFT data, we can now resum the defect-channel expansion in order to obtain the full correlation function at order $O(\veps)$:
\begin{align}
 \Gm^{\GFF}_{\frac{d-1}{3},d,v}(x,\xb)
 + \Gm_{\text{WZ}}(x,\xb)
 = \sum_{s \in -v + \mathbb{Z}} \left( 
      \sum_{n=0}^\infty \mu_{s,n} \hat f_{\Dh_{s,n}, s}(x, \xb)
    + \sum_{n=1}^\infty \mu_{s,n}^{\text{fr}} \hat f_{\Dh_{s,n}^{\text{fr}}, s}(x, \xb)
 \right) \, .
\end{align}
The GFF part can be found in equation \eqref{eq:mft-oeps} with $\delta_\phi = 1/3$.
The contribution which is new from the Wess-Zumino model is significantly harder:
{\allowdisplaybreaks
\begin{align}
\label{eq:wz-full}
 \Gm_{\text{WZ}}(x,\xb)
 & = -\frac{\veps}{3} \frac{\sqrt{x \xb}}{(1-x \xb)} \Bigg[ \nonumber \\
 & + x^v (1-v) \big(j_{2 v-1,v}(x)-j_{v,v}(x)-H_{v-1} \Phi _v(x)+\Phi _v(x) \log (x \xb)\big) \nonumber \\
 & +\xb^{1-v} (1-v) \big(j_{1-v,1-v}(\xb)-j_{2-2 v,1-v}(\xb)+H_{1-v} \Phi _{1-v}(\xb)\big) \nonumber \\ 
 & + x^v \frac{H_{v-1}-H_{2 v-2}+\Phi _v(x)-\Phi _{2 v-1}(x)}{1-x} \nonumber \\
 & + \xb^{1-v} \frac{H_{-v}-H_{1-2 v}+\Phi _{1-v}(\xb)-\Phi _{2-2 v}(\xb)}{1-\xb} \nonumber \\
 & -x^{1-v} \xb^{2-2 v} \left((v-1) J_{2-2 v,1-v}(\xb,x)+\frac{\Phi _{2-2 v}(\xb)-x \Phi _{2-2 v}(x \xb)}{1-x}\right) \nonumber \\
 & +x^{2 v-1} \xb^{v-1} \left((v-1) J_{2 v-1,v-1}(x,\xb)-\frac{\Phi _{2 v-1}(x)-\xb \Phi _{2 v-1}(x \xb)}{1-\xb}\right) \nonumber \\
 & -\xb x^{v+1} \left((v-1) J_{v+1,1}(x,\xb)-\frac{\Phi _{v+1}(x)-\xb \Phi _{v+1}(x \xb)}{1-\xb}\right) \nonumber \\
 & +x \xb^{2-v} \left((v-1) J_{2-v,1}(\xb,x)+\frac{\Phi _{2-v}(\xb)-x \Phi _{2-v}(x \xb)}{1-x}\right) \Bigg] \, .
\end{align}} 
We could not express this correlation function in terms of elementary functions.
Instead, we introduced the following two special functions
\begin{align}
\begin{split}
 j_{a,b}(x)
 \equiv \sum_{n=0}^\infty \frac{x^n H_{n+a}}{n+b} \, , \qquad
 J_{a,b}(x, \xb)
 = \sum_{n=0}^\infty \sum_{m=0}^n \frac{x^n}{(n+a)} \frac{\xb^m}{(m+b)} \, .
\end{split}
\end{align}
In appendix \ref{sec:resum-wz} we derive some interesting properties of these functions, in particular we give an efficient algorithm to generate their expansion in powers of $(1-x)$ and $(1-\xb)$.
This allows us to expand the correlation function in the bulk channel
\begin{align}
 \Gm_{\text{WZ}}(x, \xb)
 = \left( \frac{\sqrt{x \xb}}{(1-x) (1-\xb)} \right)^{(d-1)/3} \left(  
    1
    + \sum_{\ell=0}^\infty d_{\ell,n}^{\text{WZ}} F_{\Delta_{\ell,n},\ell}(x, \xb)
 \right) \, .
\end{align}
Once again, we can extract the CFT data using a software like \mathematica.
Some of the low-lying bulk OPE coefficients are
\begin{align}
\label{eq:bulk-wz-opes}
 & d_{0,0}^{\text{WZ}} 
 = \veps (v-1) \left(
 \frac{1}{3} (2 v-1) \left(H_{-2 v}+H_{2 v}\right)
 -\frac{1}{6} (3 v-2) \left(H_{-v}+H_v\right)
 -\frac{5 v^2-v+1}{6 v}\right), \nonumber \\ \quad
 & d_{1,0}^{\text{WZ}} 
 = \veps (v-1)^2 \left(
 \frac{1}{18} (2 v-1) \left(H_{-2 v}+H_{2 v}\right)
 -\frac{1}{36} (5 v-2) \left(H_{-v}+H_v\right)
 +\frac{5 v^3+10 v^2-12 v+3}{108 v (v-1)}\right), \nonumber \\ \quad
 & d_{0,1}^{\text{WZ}} 
 = \frac{\veps}{144} v (v-1) \left(17 v^2-37 v+18\right) \, .
\end{align}
Let us emphasize that the total OPE coefficients are obtained combining \eqref{eq:bulk-gff-sus-opes} and \eqref{eq:bulk-wz-opes}, namely $d_{\ell,n} = d_{\ell,n}^{\text{WZ}} + d_{\ell,n}^{\GFF}$.
An interesting feature of the CFT data is that the leading-twist family $d_{\ell,0}^{\text{WZ}}$ depends on harmonic numbers $H_a$, while all higher-twist families $d_{\ell,n\ge1}^{\text{WZ}}$ have only polynomial dependence in $v$.
This can be understood heuristically remembering that $d_\Om = \lambda_{\phi\bar\phi\Om} a_{\Om}$.
For the leading-twist family, because $\lambda_{\phi\bar\phi\Om} \sim O(\veps^0)$ then the $O(\veps)$ term in the one-point coefficient $a_\Om$ contributes to $d_\Om$.
Therefore, our result captures a one-point function calculated to one-loop in terms of Feynman diagrams, where the one-loop integrals would be responsible for the appearance of harmonic numbers.
On the other hand, for higher-twist families we have $\lambda_{\phi\bar\phi\Om} \sim O(\veps)$, so only the tree-level part of $a_\Om$ contributes to $d_\Om$, giving an intuitive reason why no harmonic numbers appear in this case.
As usual, we give a larger list of bulk coefficients in the notebook attached to this publication.

\section{Conclusions}

In this work we used analytical bootstrap techniques to study monodromy defects in the $\veps$--expansion.
This program has been highly successful for four-point functions without defects, where CFT data has been extracted up to fourth order in $\veps$ for the Wilson-Fisher fixed point \cite{Alday:2017zzv,Henriksson:2018myn}. 
Our analysis can be considered as the first step towards applying these techniques to monodromy defects in CFT.
Our main result is equation \eqref{eq:wz-full}, which describes the full leading-order two-point correlator of chiral fields in the Wess-Zumino model. In order to obtain the defect correlator, it was necessary to calculate the leading order CFT data of the Wess-Zumino model without defects (see section \ref{sec:wz-summary}), a result that is interesting on its own and that we plan to extend to higher orders in the future.

We also studied monodromy defects in the Wilson-Fisher $O(N)$ model, reproducing and in some cases improving previous results.
A natural extension of this work is to consider higher orders in the $\veps$--expansion, although this will require dealing with degeneracies in the bulk spectrum. 
Another related system is the large-$N$ limit of the $O(N)$ model, which has been studied using bootstrap in \cite{Alday:2019clp}.
Monodromy defects in the large-$N$ limit have been studied in \cite{Giombi:2021uae}, and they might be good candidates for a bootstrap analysis.

Yet another system in which the techniques used in this paper are directly applicable is a Wilson line defect in $\Nm=4$ SYM at strong coupling. 
The strong-coupling planar spectrum of $\Nm=4$ SYM contains double-trace operators which are killed by the discontinuity in the inversion formula. This is very similar to the setup of this paper, and indeed two-point functions of half-BPS operators can be reconstructed by inverting a finite number of conformal blocks \cite{Barrat:2021yvp}. It might also be possible to consider other maximally-supersymmetric models in $3 \le d \le 6$, and bootstrap their defect correlators in suitable limits. 

On a more speculative side, the functions studied in appendix \ref{sec:resum-wz} are close cousins of the Hurwitz-Lerch zeta function. Perhaps these functions will find applications in other perturbative calculations or in other branches of mathematical physics. Finally, the study of higher-point functions is one of the long-term goals of the bootstrap. Progress in this direction was made in \cite{Buric:2020zea}, where higher-point functions in the presence of defects were studied. Eventually, one should be able to obtain the corresponding Lorentzian inversion formulas, and implement the multi-point bootstrap in order to obtain even more restrictive constraints.

\section*{Acknowledgments}

We are particularly grateful to J.~Barrat, E.~Lauria and P.~van Vliet for discussions and collaboration on related projects.
AG wants to acknowledge S.~Lacroix for many useful comments.
We also thank I.~Buric, A.~Kaviraj, J.~Rong and V.~Schomerus for interesting discussions, and the anonymous JHEP referee for many comments that helped improve this work.
Finally, we thank the Simons Collaboration on the Non-perturbative Bootstrap for many stimulating activities. 
This work is supported by the DFG through the Emmy Noether research group ``The Conformal Bootstrap Program'' project number 400570283.

\appendix

\section{Appendix}

\subsection{Inverting the Wess-Zumino model}
\label{sec:app-inversion-wz}

In this appendix we explain how to obtain the Wess-Zumino defect spectrum from the discontinuity of the correlator.
By means of the inversion formula, it boils down to computing the integral \eqref{eq:inv-formula}.
An important observation is that since the discontinuity is not symmetric under $x \leftrightarrow \xb$, the integrals are different for $s>0$ and $s<0$.

Let us focus on $s>0$ first, and we summarize $s<0$ at the end.
Since the discontinuity is of order $O(\veps)$, we can evaluate the integration kernel at $d = 4$, when the integral is dramatically simpler:
\begin{align}
\label{eq:inv-int-posS}
 \mu(\Dh,s)
 = \veps \frac{v(v-1)}{12}
 \int_0^1 dx \int_1^{1/x} d\xb \, x^{-(\Dh-s+1)/2} & \xb^{-(\Dh+s+1)/2}
  \left(h\left(\tfrac{\xb-1}{\xb}\right)-h(1-x)\right) \, .
\end{align}
Let us remind the reader that $h(z) = z \, _3F_2(1,1,v+1;2,3;z)$.
The strategy to obtain the CFT data from such an integral is to notice that poles in $\Dh$ come from the region $x \to 0$.
Thus, we expand the integrand in powers of $x$ and for each power we have
\begin{align}
\label{eq:single-pow-inv}
 \int_0^1 dx \int_1^{1/x} d\xb \, x^{-(\Dh-s+1)/2} & \xb^{-(\Dh+s+1)/2} x^n
 = -\frac{2}{(s+n)} \frac{1}{(\Dh-s-1-2 n)} \, .
\end{align}
Physically, each power $x^n$ generates a defect family of dimensions $\Dh_{s,n} = 1 + s + 2 n$ and OPE coefficient $\mu_{s,n} \sim 1 / (s+n)$.
Notice that the function $h(1-x)$ has the following expansion
\begin{align}
\begin{split}
 h(1-x)
 & =
    \frac{2}{v}(1-H_{1-v})
 + \frac{2}{v(v-1)} \sum_{n=1}^\infty \left(
     \frac{(n+v-1) x^n}{n}
   + \frac{n x^{n+1-v}}{(v-n-1)} \right) \, .
\end{split}
\end{align}
Combining this expansion with the inversion \eqref{eq:single-pow-inv} one obtains the CFT data for $n > 0$, see \eqref{eq:wz-def-norm} and \eqref{eq:wz-def-weird}.
The case $n = 0$ is identical, except one also has to consider contributions from the following integral:
\begin{align}
\begin{split}
 \int_0^1 dx \int_1^{1/x} d\xb \, x^{-(\Dh-s+1)/2} & \xb^{-(\Dh+s+1)/2}
 h\left(\frac{\xb-1}{\xb}\right) \\
 & = -\frac{4 \big( (s-v+1) (H_{s-v+1} - H_s) +v-1 \big)}{s v(v-1) (\Dh-s-1)} \, . \\
\end{split}
\end{align}
This integral has been obtained by expanding the integrand around $\xb \to \infty$, integrating term by term, and finally resuming the resulting expression.
The final result can be checked numerically to very high precision.
 
Let us briefly outline the $s<0$ case.
The inversion integral is once again \eqref{eq:inv-int-posS} where one needs to change $x \leftrightarrow \xb$ in the integration region.
The CFT data for $n>0$ can be read off from the following expansion
\begin{align}
\begin{split}
 h\left( \frac{\xb-1}{\xb} \right)
 & = \frac{2}{v} \left(1-H_{v-1}+\log \xb\right)
   + \frac{2}{v (v-1)} \sum_{n=1}^\infty \left( 
      \frac{(v-n-1) \xb^n}{n}
    + \frac{n \xb^{n+v-1}}{n+v-1}
 \right) \, .
\end{split}
\end{align}
The presence of a $\log \xb$ term leads to the anomalous dimensions \eqref{eq:dim-lead-trans-twist}.
For the $n=0$ case, one also needs the integral
\begin{align}
\begin{split}
 \int_0^1 d\xb \int_1^{1/\xb} dx \, x^{-(\Dh-s+1)/2} & \xb^{-(\Dh+s+1)/2} h(1-x) \\
 & = \frac{4 \big( s (s-v+1) \left(H_{-s+v-1}-H_{-s}\right)+(s+1) (v-1)\big)}{s^2 v(v-1) (\Dh+s-1)} \, ,
\end{split}
\end{align}
which has been computed by expanding around $x = 0$ and integrating term by term.

\subsection{Defect-channel resummation}
\label{sec:resum-wz}

In this appendix we present some mathematical results that are useful in order to resum the defect-channel expansion of monodromy defects.

\subsubsection{Hurwitz-Lerch zeta function}
\label{sec:hurwitz-zeta}

The first function we consider is the well-known Hurwitz-Lerch zeta function:
\begin{align}
\begin{split}
 & \Phi(x, s, a)
 = \sum_{m=0}^\infty \frac{x^m}{(m + a)^s} \, .
\end{split}
\end{align}
The only case which is relevant in the present work is $s = 1$, when it has a simple expression as a hypergeometric function:
\begin{align}
 & \Phi(x, 1, a) 
 = a^{-1} 
   {}_2F_1\bigg( \! \begin{array}{c c} 1, \; a \\ a+1 \end{array}; x \bigg) \, .
\end{align}
The power of the Hurwitz-Lerch zeta function lies in the possibility of writing seemingly complicated infinite sums in terms of them.
Defect-channel expansions such as \eqref{eq:mft-oeps} or \eqref{eq:wz-full} can be resummed using the following formulas:
\begin{align}
\begin{split}
 & \sum _{n=0}^{\infty} x^n \left(H_{a+n}-H_{a-1}\right) 
 = \frac{\Phi(x, 1, a)}{1-x} \, , \\
 & \sum _{n=0}^\infty \sum _{m=0}^n \frac{x^{n} \xb^m}{n+a}
 = \sum _{n=0}^\infty x^n \Phi (x \xb,1,n+a)
 = \frac{\Phi (x,1,a) - \xb \Phi (x \xb,1, a)}{1 - \xb} \, , \\
 & \sum _{n=0}^\infty \sum _{m=0}^n \frac{x^{n} \xb^m}{m+a}
 = \frac{\Phi(x \xb, 1, a)}{1-x} \, .
\end{split}
\end{align}
For our applications, it is important to expand the Hurwitz-Lerch zeta function around $x = 1$, which allows us to extract the bulk CFT data.
Let us note the two elegant expressions
\begin{align}
\begin{split}
 \Phi(x, 1, v) 
 & = -\sum _{n=0}^{\infty} \frac{(v)_n}{n!} (1-x)^n 
      \big(\log (1-x) + H_{v+n-1}-H_n \big) \, , \\
 x^v \Phi(x, 1, v)
 & = -H_{v-1} 
 - \log (1-x) 
 +\sum _{n=1}^{\infty} \frac{(-1)^{n+1} (v-n)_n}{n^2 (n-1)!} (1-x)^n \, .
\end{split}
\end{align}

\subsubsection{One-variable function}
\label{sec:one-var-fun}

In the study of the Wess-Zumino model, we encountered sums that could not be expressed in terms of simple special functions.
The first sum we consider involves a single variable:
\begin{align}
\label{eq:def-jab}
\begin{split}
 & j_{a,b}(x)
 \equiv \sum_{n=0}^\infty \frac{x^n H_{n+a}}{n+b} \, .
\end{split}
\end{align}
It is not hard to relate $j_{a,b}(x)$ to itself after shifting $a \to a \pm 1$ and $b \to b \pm 1$.

Let us consider the case $a = 0$ separately. 
The function $j_{0,b}(x)$ can be resummed in terms of the incomplete beta function:
\begin{align}
\label{eq:def-j0b}
 j_{0,b}(x)
 = - x^{-b} \left. \frac{\partial B_x(b, a)}{\partial a} \right|_{a=0} \, .
\end{align}
In order to generate the series expansion around $x = 1$ efficiently, we note that the function satisfies the differential equation
\begin{align}
\label{eq:diff-eq-j0b}
 \partial_x (1-x) \partial_x (1-x) x^{1-b} 
 \partial_x \big( x^b j_{0,b}(x) \big) = 0 \, .
\end{align}
Making an ansatz for the series around $x = 1$
\begin{align}
 j_{0,b}(x)
 = \sum_{i=0}^\infty (1-x)^i \big( a_i + b_i \log(1-x) + c_i \log^2(1-x) \big) \, ,
\end{align}
one can fix coefficients recursively using the differential equation \eqref{eq:diff-eq-j0b}.
The initial condition can be obtained from \eqref{eq:def-j0b}
\begin{align}
 a_0 = -\frac{1}{2} \left((H_{b-1})^2+H_{b-1}^{(2)}\right) \, , \qquad
 b_0 = 0 \, , \qquad
 c_0 = \frac{1}{2} \, .
\end{align}
Here $H^{(r)}_b = \sum_{n=1}^b n^{-r}$ is a generalization of the harmonic number, where the usual continuation to non-integer values of $b$ is assumed.

Let us move on to the general case $a \notin \mathbb{N}$, and define the auxiliary function
\begin{align}
 \tilde j_{a,b}(x)
 = \sum_{n=0}^\infty \frac{x^n (H_{n+a} - H_{a-1})}{n+b} 
 = j_{a,b}(x) - H_{a-1} \Phi(x, 1, b) \, .
\end{align}
Clearly, any property of $\tilde j_{a,b}(x)$ can be easily translated to $j_{a,b}(x)$, since the Hurwitz-Lerch zeta function that relates them is well understood.
The advantage of the auxiliary function is that it satisfies a simpler differential equation
\begin{align}
  \partial_x (1-x) x^{1-a} \partial_x (1-x) x^{a+1-b} 
  \partial_x \big( x^b \tilde j_{a, b}(x) \big) = 0 \, .
\end{align}
From this differential equation, one can efficiently generate the expansion around $x = 1$ fixing the coefficients in the ansatz
\begin{align}
 j_{a,b}(x)
 = \sum_{i=0}^\infty (1-x)^i \big( d_i + e_i \log(1-x) + f_i \log^2(1-x) \big) \, .
\end{align}
In order to find the initial conditions $d_0$, $e_0$ and $f_0$, we note that the sum \eqref{eq:def-jab} can be obtained in \mathematica in terms of complicated special functions.
Taking the $x \to 1$ limit, and massaging the resulting expressions, we find
\begin{align}
\begin{split}
 d_0 & = -\sum_{n=0}^\infty \left( \frac{H_{a-b+n+2}}{n+2}-\frac{H_{a+n+2}}{b+n+2} \right) +\frac{1}{2} \left(H_{a-b}\right){}^2-H_{a-b}+H_a \left(H_b+\frac{1}{b+1}\right) \\ 
 & \; \; \; -\frac{H_b}{a}+\frac{H_{a-b}^{(2)}}{2}+\frac{1}{a b}+\frac{1}{-a+b-1}+\frac{1}{a b+a+b+1}+\frac{\pi ^2}{6} \, , \\
 e_0 & = H_{a+2}-\frac{1}{a}-\frac{1}{a+1}-\frac{1}{a+2} \, , \\
 f_0 & = \frac12 \, .
\end{split}
\end{align}
We have not been able to further simplify the infinite sum in $d_0$.
However, it is interesting to note that when expanding \eqref{eq:wz-full} in the $x,\xb \to 1$ limit, we have found numerically that the contributions from these infinite sums combine to give zero.

\subsubsection{Two-variable function}

There is another type of double sum that we have not been able to express in closed form:
\begin{align}
\begin{split}
\label{eq:def-Jab}
 & J_{a,b}(x, \xb)
 \equiv \sum_{n=0}^\infty \sum_{m=0}^n \frac{x^n}{(n+a)} \frac{\xb^m}{(m+b)} \, .
\end{split}
\end{align}
For the bulk channel expansion, we need the series expansion of $J_{a,b}(x,\xb)$ around $x,\xb = 1$.
For simplicity, we always take the limits in the order $|1-x| \ll |1-\xb|$.
There is no loss of generality, since in order to expand the function $J_{a,b}(\xb,x)$, one can use the relation
\begin{align}
\begin{split}
 & x J_{a,b}(x,\xb)
 = \Phi (x,1,a-1) \Phi (\xb,1,b)-J_{b,a-1}(\xb,x) \, , \\
\end{split}
\end{align}
which follows from the definition \eqref{eq:def-Jab}.
The strategy to expand around around $x,\xb = 1$ is to first compute the sum in $x$,
and then expand only in $x \to 1$:
\begin{align}
 J_{a,b}(x, \xb)
 & = \sum_{m=0}^\infty \frac{(x \xb)^m}{m+b} \Phi(x, 1, a+m) \\
 & = - \sum _{n=0}^{\infty} (1-x)^n \sum _{m=0}^{\infty} \sum _{p=0}^n 
   \frac{\xb^m  (-m)_p (a+m)_{n-p} \left(H_{a+m+n-p-1}-H_{n-p}+\log (1-x)\right)}{p! (b+m) (n-p)!} \, . \nonumber
\end{align}
Now we perform that sum in $(1-x)^n$ to the desired order $n_{\text{max}}$.
For any finite value of $n_{\text{max}}$, we compute the finite sum in $p$, and then the sum in $m$ can be computed in terms of rational functions of $(1-\xb)$ and the function $j_{a,b}(\xb)$.
Using the results of section \ref{sec:one-var-fun}, we finally obtain the expansion in $(1-x)$ and $(1-\xb)$ to any desired order.
Although it would be hard to do this by hand, the previous algorithm can be implemented efficiently in \mathematica.
Let us also note that the series expansion contains terms of the form $(1-x)^n (1-\xb)^{-m}$ for $n,m \ge 0$.
A good sanity check of our implementation is that these spurious powers cancel when they are combined as in \eqref{eq:wz-full}.


\providecommand{\href}[2]{#2}\begingroup\raggedright\endgroup

\end{document}